\documentclass[aps,twocolumn,prd,superscriptaddress,showpacs,nofootinbib,fixfloat]{revtex4}
\usepackage{graphicx}
\usepackage{dcolumn}
\usepackage{bm}
\usepackage{natbib}

\newcommand{\be}{\begin{equation}}
\newcommand{\ee}{\end{equation}}
\newcommand{\bea}{\begin{eqnarray}}
\newcommand{\eea}{\end{eqnarray}}

\newcommand{\Fermi}{{\slshape Fermi}}

\newcommand{\s}{{\rm ~s}}

\newcommand{\cm}{{\rm ~cm}}
\newcommand{\ph}{{\rm ~ph}}
\newcommand{\sr}{{\rm ~sr}}

\newcommand{\Hz}{{\rm ~Hz}}

\newcommand{\GeV}{{\rm ~GeV}}

\newcommand{\degree}{^{\rm o}}

\newcommand{\zrock}{$Z_{\rm rock}$}

\def\la{\vcenter{\hbox{$<$}\offinterlineskip\hbox{$\sim$}}}

\newcommand\Refsec[1]{Section \ref{sec:#1}}

\begin{document}

\title{Is the 130 GeV Line Real? 
  A Search for Systematics in the Fermi-LAT Data}

\author{Douglas P. Finkbeiner}
\affiliation{Institute for Theory and Computation,
  Harvard-Smithsonian Center for Astrophysics, 
  60 Garden Street, MS-51, Cambridge, MA 02138, USA} 
\affiliation{Center for the Fundamental Laws of Nature,
  Physics Department, 
  Harvard University, 
  Cambridge, MA 02138 USA}

\author{Meng Su}
\affiliation{Institute for Theory and Computation,
  Harvard-Smithsonian Center for Astrophysics, 
  60 Garden Street, MS-51, Cambridge, MA 02138, USA} 
\affiliation{Department of Physics, and Kavli Institute for Astrophysics and Space Research, Massachusetts Institute of Technology, Cambridge, MA 02139, USA}
\affiliation{Einstein Fellow}

\author{Christoph Weniger}
\affiliation{Max-Planck-Institut f\"ur Physik, 
  F\"ohringer Ring 6, 
  80805 M\"unchen, Germany}

\begin{abstract} Our recent claims of a Galactic center
  feature in \Fermi-LAT data at approximately 130 GeV have
  prompted an avalanche of papers proposing explanations
  ranging from dark matter annihilation to exotic pulsar
  winds.  Because of the importance of such interpretations
  for physics and astrophysics, a discovery will require not only additional
  data, but a thorough investigation of possible LAT
  systematics.  While we do not have access to the details
  of each event reconstruction, we do have information about
  each event from the public event lists and spacecraft
  parameter files.  These data allow us to search for
  suspicious trends that could indicate a spurious signal.
  We consider several hypotheses that might make an instrumental
  artifact more apparent at the Galactic center, and find them 
  implausible.  We also search for an instrumental signature in 
  the Earth limb photons, which provide a smooth
  reference spectrum for null tests.  We find no significant 
  130 GeV feature in the Earth limb sample.  However, we do 
  find a marginally significant 130 GeV feature in Earth 
  limb photons with a limited range of detector incidence angles.  
  This raises concerns about
  the 130 GeV Galactic center feature, even though we can
  think of no plausible model of instrumental behavior that
  connects the two.  A modest amount of additional limb data
  would tell us if the limb feature is a statistical fluke.
  If the limb feature persists, it would raise doubts
  about the \texttt{Pass 7} processing of $E > 100$ GeV events.  At present
  we find no instrumental systematics that could plausibly explain the excess
  Galactic center emission at 130 GeV. 
\end{abstract}


\maketitle



\section{Introduction}

The search for non-gravitational signatures from WIMP
(weakly interacting massive particle) dark matter has 
generally been approached from three different directions: missing
energy searches at colliders, direct searches for the
recoil of nuclei from underground detectors, and indirect
methods including searching for dark matter signals from cosmic
rays (CR) and multiwavelength astronomical
observations~\citep{Jungman:1995df, Bergstrom:2000, Bertone:2005, Hooper:2007Review,
2012arXiv1205.4882B, Cirelli:2012tf}.

For indirect detection, distinguishing the dark matter
signal from conventional astrophysical backgrounds is
challenging
(for a recent review on indirect searches with gamma rays
see~\cite{Bringmann:2012ez}).
Among various possible signatures, gamma-ray
line emission is a long-sought ``smoking
gun'' for dark matter annihilation~\cite{Bergstrom:1988fp}, as no plausible
astrophysical background can produce such a line
signature.\footnote{A narrow feature is
possible in theory~\citep[see][]{2012arXiv1207.0458A}.}  Gamma-ray line(s)
could be produced by dark matter decays or annihilations
into two photons, or two-body final states involving one
photon plus a Higgs boson, Z boson, or other neutral non-SM
particle.  In most models, the branching ratio
to lines is loop suppressed relative to the continuum
emission, and one would have expected to see the continuum
first in e.g. MSSM models~\citep[e.g.][]{Bergstrom:1997}.
Although this theoretical prejudice led most previous
studies to focus on continuum searches, there are models
being proposed that allow high line to continuum
ratios~\citep[e.g.][]{Bergstrom:1998, Bergstrom:2000,
Bertone:2009, Jackson:2010, Cline:2012, Weiner:2012}.
However, previous searches in EGRET~\cite{Pullen:2006sy} and \Fermi-LAT
data~\cite{Abdo:2010nc, Vertongen:2011mu, Ackermann:2012qk}
did not find any
indications for a gamma-ray line signal and presented only upper limits on the
line flux.

First indications for a spectral feature around 130 GeV were found by
Bringmann \textit{et al.}~\citep{Bringmann:2012} in context of virtual
internal Bremsstrahlung signals from annihilations. The first claim for a
significant line at the Galactic center (GC) was made by
Weniger~\citep{Weniger:2012}.
Both works
focused on spectral fitting to photon events in regions of interest in
the inner Galaxy designed to maximize S/N. Weniger found a line structure
with 4.6$\sigma$ (3.2$\sigma$ after the trials factor
correction) at 130 GeV, and argued against an obvious instrumental cause.
This claim was quickly followed up and disputed by a number of
groups~\cite{tempel:2012ey, Boyarsky:2012ca}.

Subsequent work by Su \& Finkbeiner approached the problem
with template fitting, which takes into account the spatial
distribution of events along with spectral information,
assuming various profiles (Einasto, NFW, Gaussian) for the
DM distribution~\citep{linepaper}.  If the template is
correct, this allows extraction of the DM signal with higher
S/N.  This work found 6.6$\sigma$ (5.1$\sigma$ after the
trials factor correction) for an Einasto profile centered
$1.5\degree$ west of the Galactic center, and also suggested
that there may be two lines, at about 111 and 129 GeV.  The
lower energy line is tantalizing because it matches the
expected energy of a $Z\gamma$ line if the higher energy is
the $\gamma\gamma$ line.  These findings have inspired a
number of models and further analysis of the \Fermi\
data~\citep{Dudas:2012, Choi:2012, Kyae:2012, Lee:2012,
Rajaraman:2012, Acharya:2012, Garny:2012, Buckley:2012,
Chu:2012, Kang:2012, Buchmuller:2012, Bergstrom:2012b,
Heo:2012, Park:2012, Tulin:2012, Cline:2012, Weiner:2012,
WeinerYavin:2012b, FanReece:2012, Huang:2012, Whiteson:2012,
Buchmuller:2012, Cholis:2012}.

Recent evidence for lines at 111 GeV and 129 GeV with a
local significance of $3.3\sigma$ from \Fermi\ unassociated
point sources suggests an annihilation signal is
present~\cite{doubleline}\citep[but
see][]{HooperLinden:2012b}, as does the claim of line
emission from galaxy clusters at 130
GeV~\cite{Hektor:2012kc}.  Neither of these would stand on
their own, but they provide support for the hypothesis that
the Galactic center line signal is produced by dark matter
annihilation.

The high statistical significance of the line feature
motivates a search for systematic errors in the LAT data
that could mimic a line in the Galactic center.
Confirmation by Imaging Air Cherenkov Telescopes like
HESS-II might be possible as early as next
year~\cite{Bergstrom:2012}, but in the meantime a thorough
study of LAT systematics is urgently needed.  We do not have
access to the details of the reconstruction of each photon
event, which would allow us to study how it developed in the
tracker and calorimeter.  However, we do have information
about each event from the public event lists and spacecraft
parameter files.  We can use this information to search for
any line-producing artifacts in the detector frame, and
investigate if they could map onto the Galactic center.

The Earth's atmosphere provides a convenient source of
photons for systematics tests.  The continual cosmic-ray
cascades in the Earth's atmosphere produce gamma rays with
$dN/dE \sim E^{-2.8}$~\citep{FermiLimb}.  Because these
so-called `Earth limb photons' result from atmospheric
cascades, they are produced by interactions in a highly
boosted frame, and cannot contain line emission.
\medskip

The rest of the paper is organized as follows: In \Refsec{130GeV} we briefly
define key parameters of LAT photon events and describe the survey strategy.
We examine peculiarities of the Galactic center observation, possible
systematics of the LAT in the instrumental frame, and study whether they could
fake a 130 GeV line signature. In \Refsec{EarthLimb}, we concentrate on a
suspicious subset of Earth limb photons that shows a line-like excess at 130
GeV. We search for correlations with the GC line events, and introduce an
energy remapping model as possible explanation for spurious signals.  Finally in
\Refsec{Conclusion}, we discuss our findings and what is required to clarify
the status of the 130 GeV excess.


\section{The 130 GeV excess}
\label{sec:130GeV}
In this section, we briefly summarize the standard survey strategy of \Fermi\ 
and define the basic parameters of each event.  We
address the question of whether observations of the Galactic center
are peculiar in a way that could enhance instrumental
effects towards this direction and search for suspicious trends in
other regions of the sky and the reconstruction parameters.


\begin{table*}
  \begin{center}
    \begin{tabular}{lcrrcl}
      \hline
      Parameter &\hphantom{i}& & Range &\hphantom{i}&  Description\\
      name      && min & max &&            \\
      \hline
      $\theta$ &&    0 &  $\sim80$ && Polar coordinate (instrument frame) \\
      $\phi$   &&    0 &       360 && Azimuthal coordinate wrt $+x$ (instrument frame) \\
      $(x,y,z)$&&  $-1$& $1$ && Cartesian coordinates;
      $\propto(\sin\theta\cos\phi, \sin\theta\sin\phi, \cos\theta)$\\
      $Z$      &&    0 & $\sim113$ && Zenith angle (horizon at $Z=113\degree$) \\
      $\ell$   && -180 & 180 && Galactic longitude \\
      $b$      &&  -90 &  90 && Galactic latitude \\
      $\psi$   &&    0 & 180 && Angle to Galactic center; $\cos\psi\equiv\cos\ell\cos b$ \\
      \zrock\  && -110 & 110 && Rocking angle (boresight angle N of zenith) \\
      \hline
    \end{tabular}
    \caption{Event parameter definitions. The Cartesian coordinates $(x, y, z)$ are
    defined such that the $z$-axis corresponds to the LAT boresight, and the
    $y$-axis is parallel to the solar panels.}
    \label{tab:parameters}
  \end{center}
\end{table*}

\subsection{Standard survey strategy and definitions}
\label{sec:conventions}









With a field of view of $\sim2.5$~sr, the LAT can survey the
entire sky in two orbits.  In \emph{standard survey mode},
the LAT points north of zenith towards the orbital pole by
an angle \zrock\ on one orbit, and south of zenith by the
same angle on the next orbit.  In this mode, the LAT
pointing is confined to the plane perpendicular to its
orbital velocity.  The slews are performed with a repeating
pattern of 17 waypoints defining \zrock\ as a function of
time.\footnote{The survey rocking angle profiles are
available at
\url{http://fermi.gsfc.nasa.gov/ssc/observations/types/allsky/}.
The effective dates and times for each profile are provided.
} \zrock\ was $35\degree$ at the start of the nominal
mission on August 4, 2008 until May 7th, 2009, and was
changed to $50\degree$ on September 3rd, 2009 for better
thermal management of the downward-facing battery
radiator.\footnote{Various profiles with rocking angle
39$\degree$, 40$\degree$, and 45$\degree$ respectively have
been tested for relatively short periods, including a 3
orbit profile test that overweights the south.}  This
rocking-angle profile, combined with the precession of the
orbit every $\sim53.4$ days, allows the LAT to observe the
whole sky with approximately uniform coverage.

\Fermi\ spends over 95\% of the mission time in standard survey mode.
This is only occasionally interrupted for
pointed observations of targets of opportunity (ToOs).  During such times
the LAT may point at larger zenith angle than usual, even at the horizon.
\Fermi's survey observations are
halted during passages through the South
Atlantic Anomaly, resulting in an exposure
differential between north and south of $\sim15$\%. In addition,
survey mode is occasionally interrupted by Autonomous
Repoints of the observatory for triggered
gamma-ray burst follow-up observations, and for calibration.
\medskip

The reconstructed arrival direction of photons in celestial
coordinates, LAT coordinates, and Earth coordinates is
described by parameters in Table~\ref{tab:parameters}.
$\theta$ is the reconstructed incidence angle of the photon
event with respect to the LAT boresight (defined as the $+z$
axis).  The $+x$ axis is the line normal to the Sun-facing
side of the spacecraft, i.e.~the solar panels, which are
parallel to the $y$ axis, face roughly the $+x$ direction.
$\phi$ is the azimuthal angle of incidence with respect to
the $+x$ axis. The Zenith angle $Z$ is the angle between the
reconstructed event direction and the zenith line, which
passes from the Earth center through the satellite center.
All angles are in units of degrees.

Due to the increased rocking angle \zrock$\simeq50\degree$
since September 2009, photons from the Earth limb
entered the FOV of the LAT:
At a spacecraft altitude $a$, the geometric (unrefracted) horizon is seen at zenith angle \be
Z_{\rm hor} = \cos^{-1}\left(\frac{R_\oplus}{R_\oplus+a}\right)+90\degree\;. \ee
The \Fermi\ orbit is nearly circular with 535~km $< a <$ 564~km, yielding
$Z_{\rm hor}$ in the 112.7$\degree$ to 113.3$\degree$ range, with the tangent
point some 2400 km distant.  At this distance, the $\sim 100$ km height of the
atmosphere subtends about $2\degree$, or roughly $111\degree < Z <
113\degree$. Combined with the large rocking angle, the Earth limb events are
dominantly seen near the incidence angle $\theta=Z_{\rm hor}-Z_{\rm rock}\sim 62^\circ$.
\medskip

\begin{table*}
  \begin{tabular}{lllrrr}
    \hline
    Sample &&Cuts & $N(>100\GeV)$ & $\frac{N(>100\GeV)}{N(>30\GeV)}$ & $\frac{N(>300\GeV)}{N(>100\GeV)}$\\
    \hline
    Standard events      &  & $Z<100^\circ$ & 5093 & 13.4\% & 9.6\% \\
    Inner Galactic plane &  & $Z<100^\circ$, $3^\circ < |\ell| < 30^\circ,\ |b|<2^\circ$     & 703 & 16.9\% & 9.8\% \\
    Galactic center      &  & $Z<100^\circ$, $\psi<3^\circ$ & 82 & 17.4\% & 9.8\% \\
    Galactic center line &  & $Z<100^\circ$, $\psi<3^\circ$, $120\GeV<E<138\GeV$             & 26 & -- & -- \\
    Earth limb           &  & $Z>110^\circ$ & 3120 & 10.2\% & 9.2\% \\
    Earth limb line      &  & $Z>110^\circ$, $30^\circ<\theta<45^\circ$, $120\GeV<E<138\GeV$ & 45 & -- & -- \\ 
    \hline
  \end{tabular}
  \caption{Definition of six samples of events used throughout this work.  The
    number with $E>100$ GeV is given for each sample, along with the indicated
    ratios.  All samples have $10\leq E\leq500\GeV$.}
  \label{tab:regions}
\end{table*}

In Table~\ref{tab:regions} we define the event samples
used throughout this work: `Standard events' are all
events with the zenith angle cut $Z<100^\circ$ recommended
by the LAT team to exclude Earth
limb photons (see Section~\ref{sec:EarthLimb}); `Inner
Galactic plane' refers to a part of the Galactic disk close
to but without the center; `Galactic center' events come from
a radius of $3^\circ$ around
$\ell=b=0^\circ$; `Earth limb events' have
zenith angles $Z>110^\circ$, and are completely
dominated by photons generated in CR cascades in the
atmosphere. `Line' events refer to subsets with energies
between $120$ and $138$ GeV. This energy range is selected
since the dominant line at the GC is found to be around 129
GeV~\cite{linepaper}, and the FWHM of the relevant LAT energy dispersion is about
13.6\% at that energy~\cite{Weniger:2012}.

Throughout, we will use \texttt{P7CLEAN\_V6} events from
Aug 4th 2008 to September 5th 2012, with $10\GeV\leq E\leq 500 \GeV$ and the good-time-interval cuts
\texttt{DATA\_QUAL==1} and \texttt{LAT\_CONFIG==1}, however
without the commonly adopted cut on the rocking angle
$|$\zrock$|<52^\circ$, unless otherwise stated. This last cut would remove
low incidence angle Earth limb events, which will be of special interest
below.
\medskip

\begin{figure}[h]
  \begin{center}
    \includegraphics[width=0.99\linewidth]{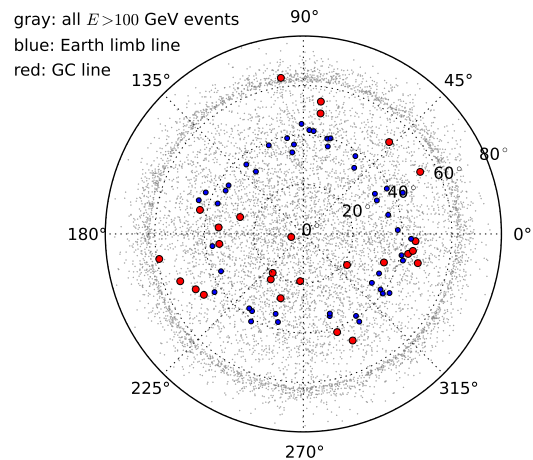}
    \vspace{-0.5cm}
  \end{center}
  \caption{Incidence angle distribution of \emph{all} events
  with $E>100\GeV$ (gray dots; see
  discussion in Sec.~\ref{sec:conventions}), of GC line
  events (red dots; see Table~\ref{tab:regions}) and of
  suspicious Earth limb events (blue dots; see
  Table~\ref{tab:regions} `Earth limb line'), as a function of
  the instrumental coordinates $\theta$ and $\phi$. The
  Earth limb contribution is clearly visible in the gray
  dots at $\theta > 60^\circ$.  Note that just because a GC line event was
  observed at high $\theta$ \emph{does not} mean it was observed near the
  horizon.}
  \label{fig:phiThetaDist}
\end{figure}

In
Fig.~\ref{fig:phiThetaDist}, the gray dots show the
distribution of \emph{all} events (i.e.~without the $Z$ cut)
with energies above 100 GeV as a function of the instrumental
incidence angles $\theta$ and $\phi$.
The contribution from the Earth limb is clearly visible at
$\theta>60^\circ$. For comparison, the red and blue dots
show the `GC line' and the `Earth limb line' events that
will be discussed below.

\subsection{Peculiarities of the Galactic center observation}
The fact that the dominant 130 GeV line signal is near the
Galactic center raises a number of concerns.  The gamma-ray
flux at the GC is somewhat brighter and might have a harder
spectrum than neighboring regions. Also, the GC is near the
ecliptic ($\beta \approx -5\degree$) and is observed in
a restricted range of angles in instrument coordinates
$(\theta, \phi)$ when the Sun passes near it.  We consider
whether these facts could exacerbate any systematic errors
in the LAT data to produce a spurious signal.

\subsubsection{Hypothesis: The Galactic center is bright, so
instrumental artifacts are more significant there.}

\begin{figure}
  \centering
  \includegraphics[width=1.0\linewidth]{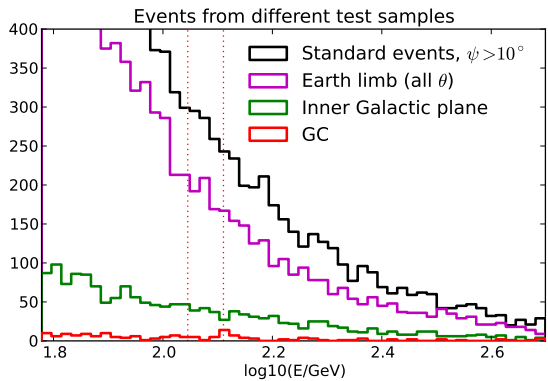}
  \caption{Energy distribution in various event samples 
      as
  discussed in the text. None of them show the excess events
  around 130 GeV seen in the Galactic center. The red dotted
  lines indicate 111 and 129 GeV.}
  \label{fig:target_spectra}
\end{figure}

The hardware trigger rate of the LAT is typically
about $10^3$--$10^4\Hz$, with the rate of
accepted \texttt{SOURCE} and \texttt{CLEAN} class events below
$3\Hz$~\citep{collaboration:2012kca}, and the rate of $E>100\GeV$ events
orders of magnitude lower. In light of these low trigger rates
(and assuming steady sources and CR background fluxes) the
LAT instrumental response cannot
depend on the brightness of an observed region.
Furthermore, at $E > 100$ GeV, the Galactic center is only modestly brighter
than the surrounding regions, so that related effects should also appear away
from the GC. Otherwise, fake 130 GeV events would have to be mistakenly
mapped from either lower energy ($E \la 10$ GeV) gammas or much lower energy
photons (e.g.  X-rays from the 1E 1740.7-2942 microquasar~\cite{Gallo:2002} or
511 keV photons~\cite{Prantzos:2011}) in which the Galactic center is much
brighter.  It is difficult to see how this could happen.

Bright regions provide samples with a high gamma-ray-to-CR ratio,
which are used for calibration purposes by the LAT team~\cite{collaboration:2012kca}.
Besides that, their main virtue is that they feature a large number of events so
that the impact of an instrumental effect, like e.g.~energy
reconstruction or acceptance anomalies that affect the reconstruction of
gamma-ray events, can be statistically more
significant there. 
We check as a warm-up whether we find indications for
suspicious features at 130 GeV.
Besides the Galactic center (with an intensity of $\sim 8\times10^{-8}
\ph\cm^{-2}\s^{-1}\sr^{-1}$ above 100 GeV), we consider the
Earth limb (with \emph{all} incidence angles and an intensity of $\sim3\times10^{-7}
\ph \cm^{-2}\s^{-1}\sr^{-1}$ above 100~GeV~\cite{FermiLimb}) and the inner
Galactic plane
($\sim9\times10^{-8}\ph\cm^{-2}\s^{-1}\sr^{-1}$), see
Table~\ref{tab:regions}. 
The sample with the highest number of celestial photons 
(but a smaller intensity, $\sim5\times10^{-9}\ph\cm^{-2}\s^{-1}\sr^{-1}$, and hence
a higher CR contamination) is
simply the whole sample of
standard events excluding the region around the GC
($\psi>10^\circ$).
For these samples, we show the corresponding energy
distributions in Fig.~\ref{fig:target_spectra}. None of the
samples exhibit an $\mathcal{O}(1)$ excess at 130 GeV as
observed in the Galactic center. Note that in the vicinity
of the bright point sources Vela and Geminga (within
$0.8^\circ$ radius, corresponding to the $95\%$ containment angle) 
only four $E>100\GeV$ events were measured, at
102.5, 111.5, 123.8 and 205.5 GeV.
\medskip

\begin{figure}
  \centering
  \includegraphics[width=0.48\textwidth]{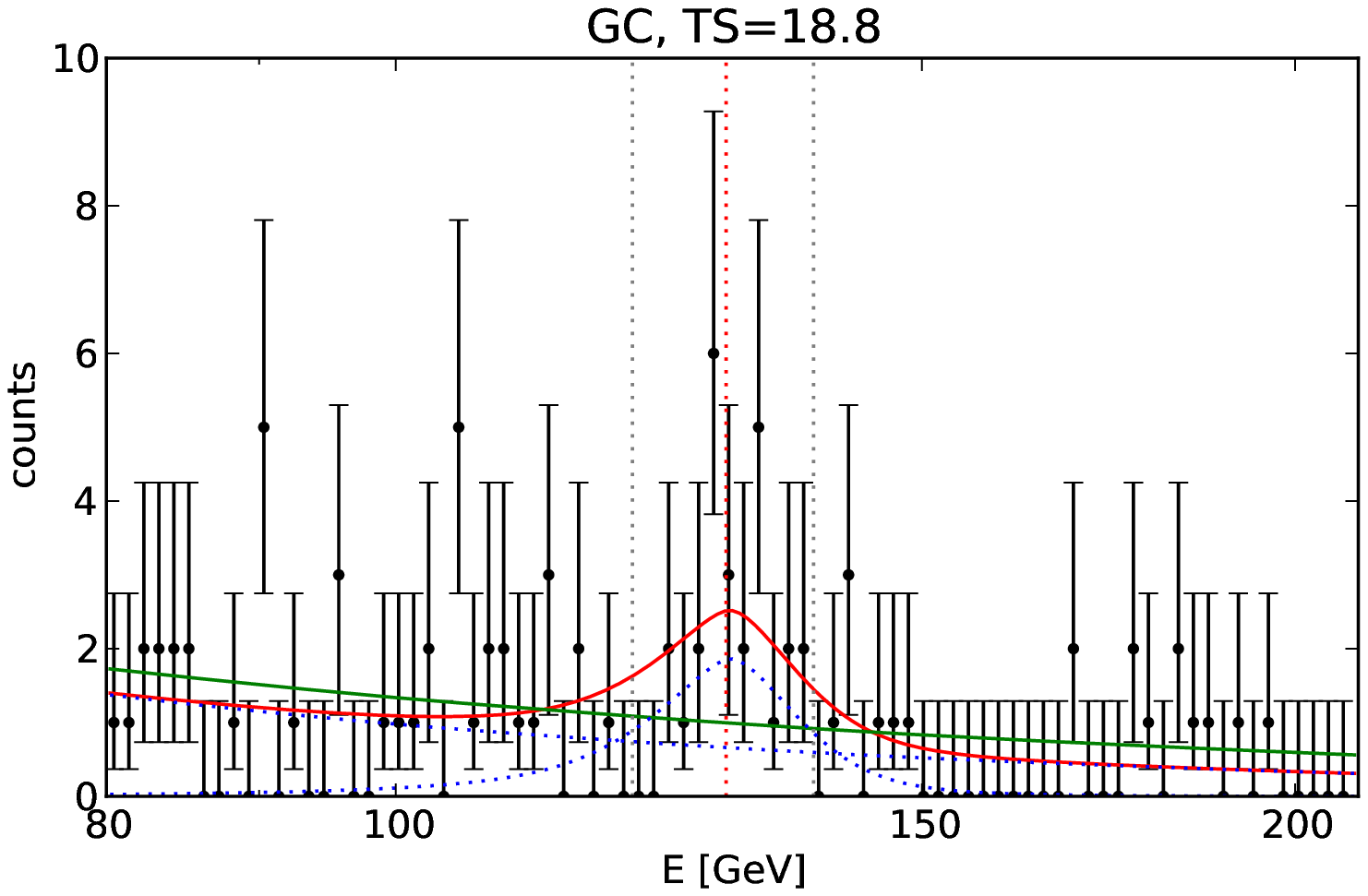}
  \includegraphics[width=0.48\textwidth]{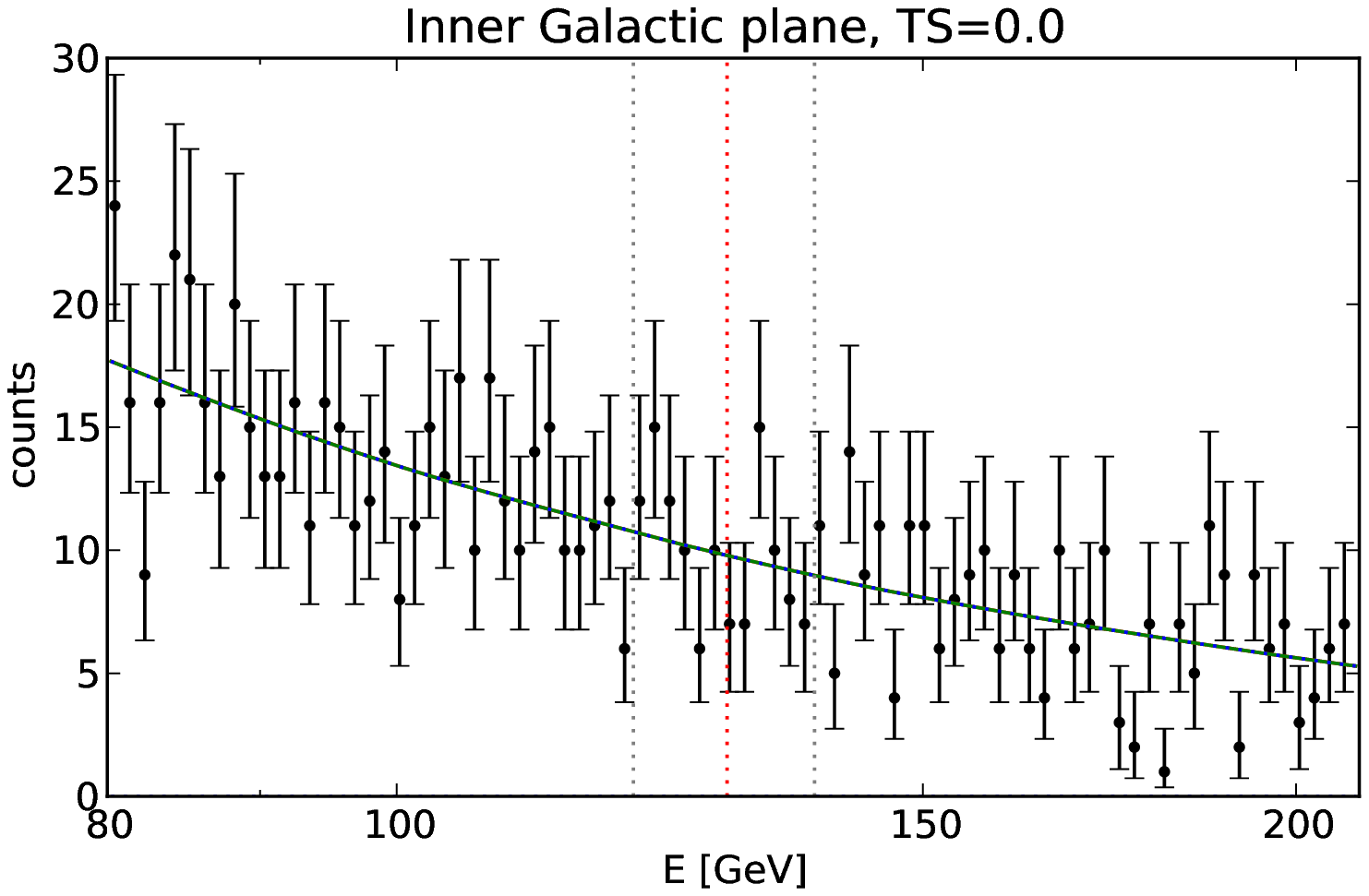}
  \includegraphics[width=0.48\textwidth]{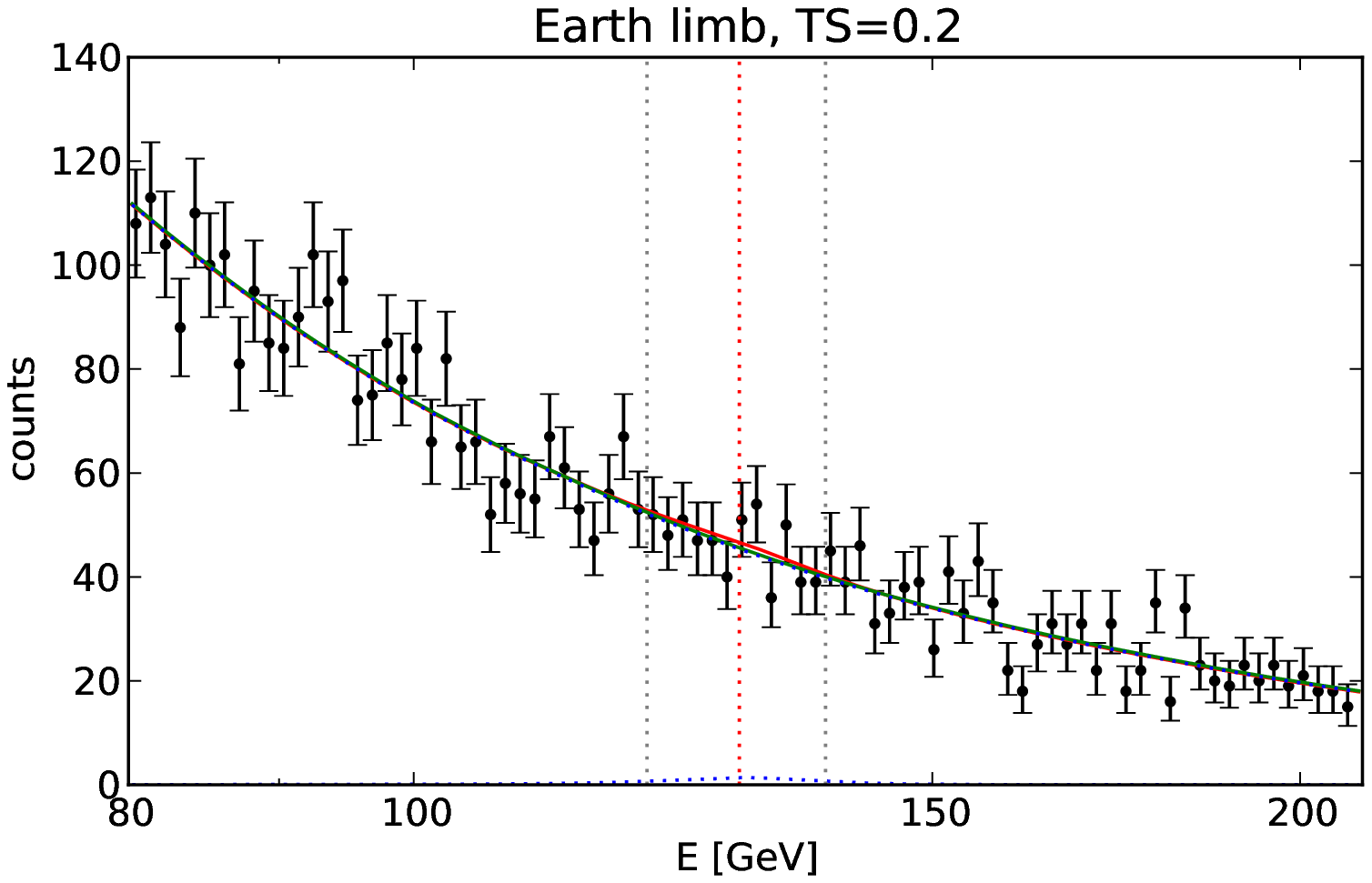}
  \caption{Spectral fits to the GC, inner Galactic plane, and Earth limb
    samples. The green line
  shows the null model (a power-law), whereas the red line
  shows the alternative power-law + line fit; the dotted
  blue lines are the two components of the alternative
  model. The red (black) dotted lines indicate 129 GeV
  (13.6\% FWHM around 129 GeV). 
  Note that the significance found in the GC region does not represent
  the full significance of the putative signal.}
  \label{fig:spectra1}
\end{figure}

In Fig.~\ref{fig:spectra1} we show fits to the energy
spectra of the GC, inner Galactic plane, and Earth limb.  The model fits a
line on a power-law background, convolved with the instrumental response, as
in~\cite{Weniger:2012}. The $TS$ value
is defined as $TS=-2\ln \mathcal{L}_{\rm
null}/\mathcal{L}_{\rm alt}$, where $\mathcal{L}_{\rm
null(alt)}$ is the likelihood of a fit without(with) a line.  The line
normalization is constrained to be non-negative. While the TS value of 18.8 in
the GC is relatively large, there is no indication for line contributions in
the Inner Galactic plane or Earth limb samples.

We emphasize that signal significances calculated from the GC region as
defined in this paper \emph{do not represent the full significance of the
  putative signal}, which is higher in regions with optimized signal-to-noise
ratio~\cite{Bringmann:2012, Weniger:2012} or when extracted by a template
analysis~\cite{linepaper}. We use the GC region as defined here since it
should be dominated by line events, making it a good starting point to look
for suspicious trends in the data.

%

\begin{figure*}
  \centering
  \includegraphics[width=0.44\textwidth]{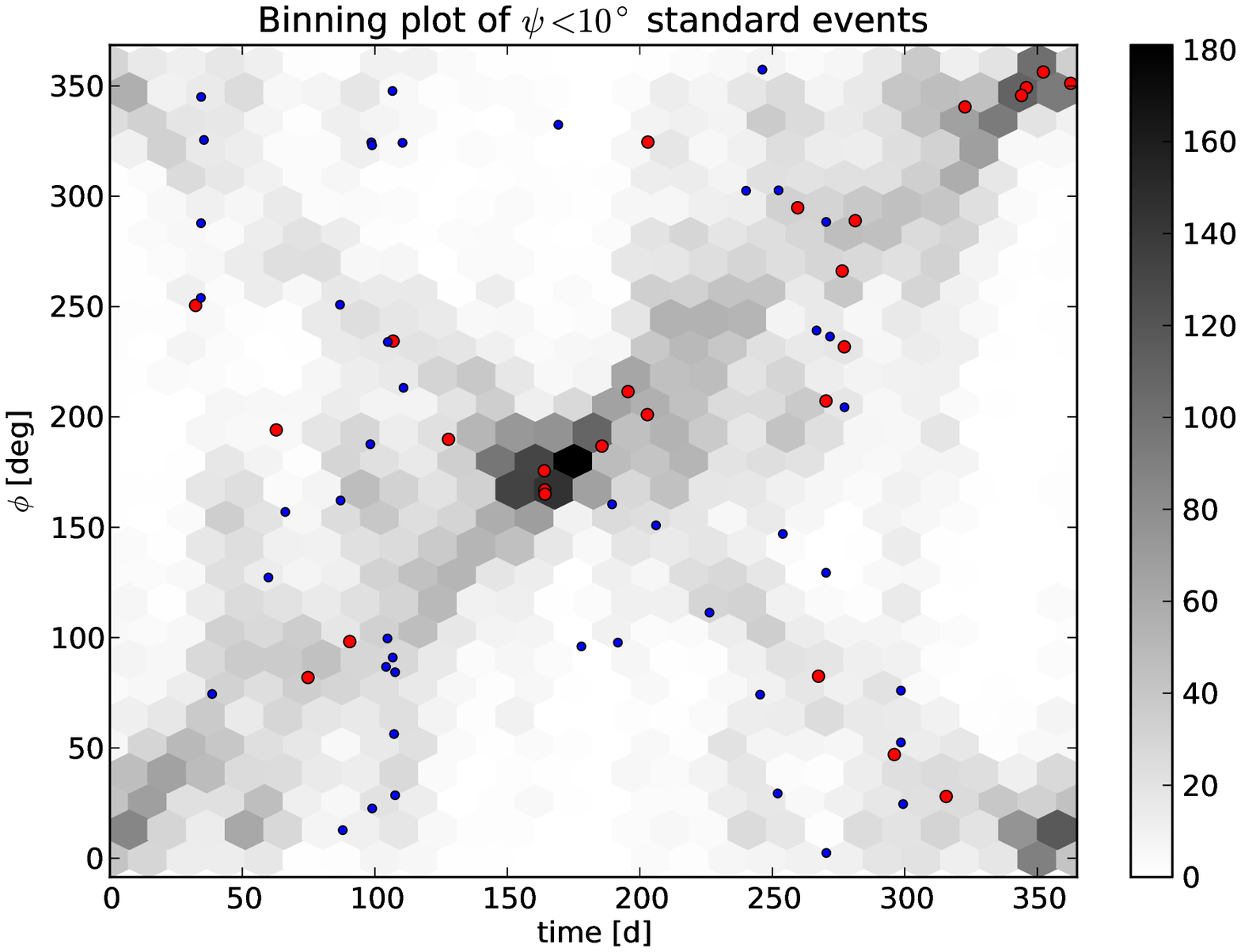}
  \includegraphics[width=0.44\textwidth]{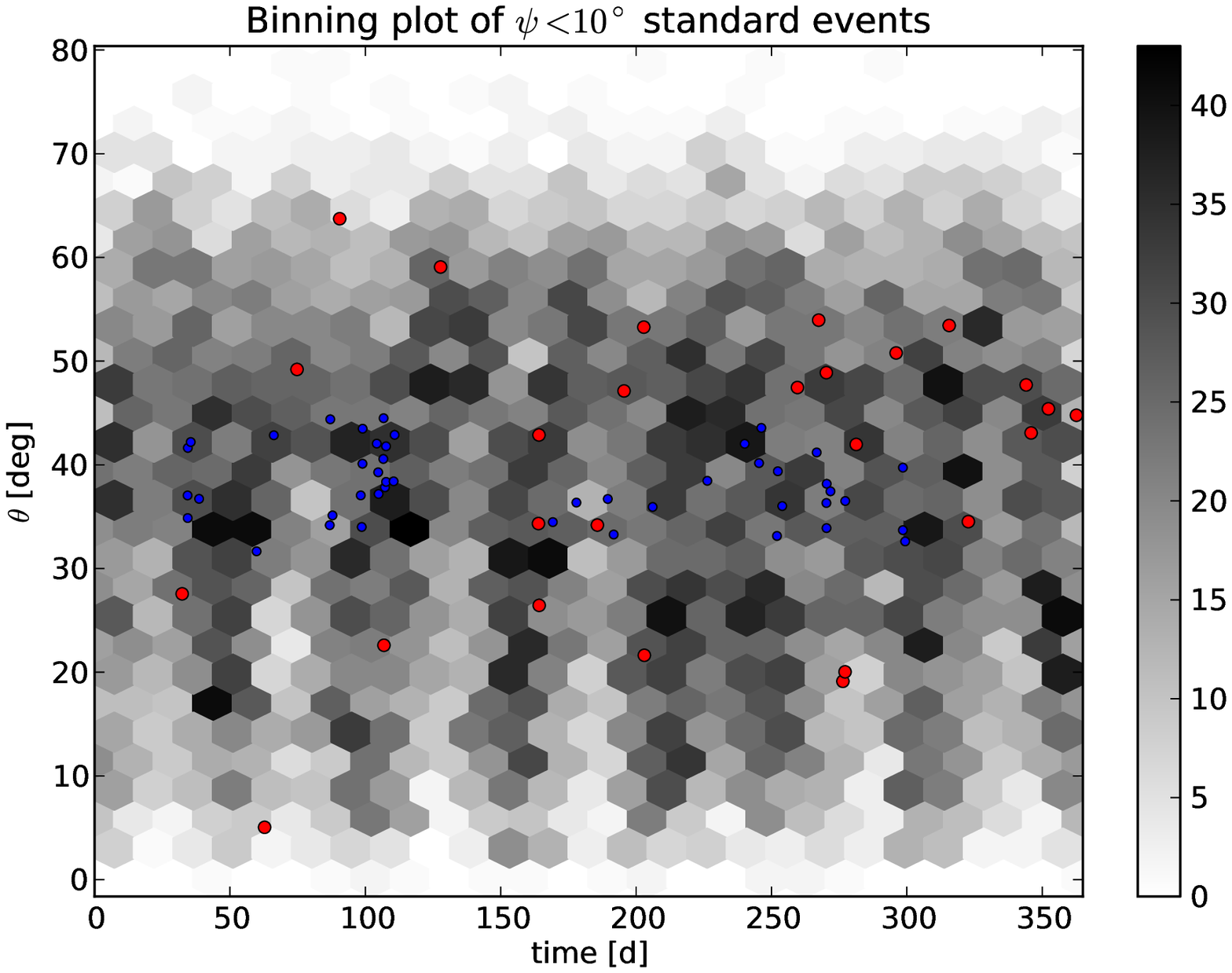}
  \includegraphics[width=0.44\textwidth]{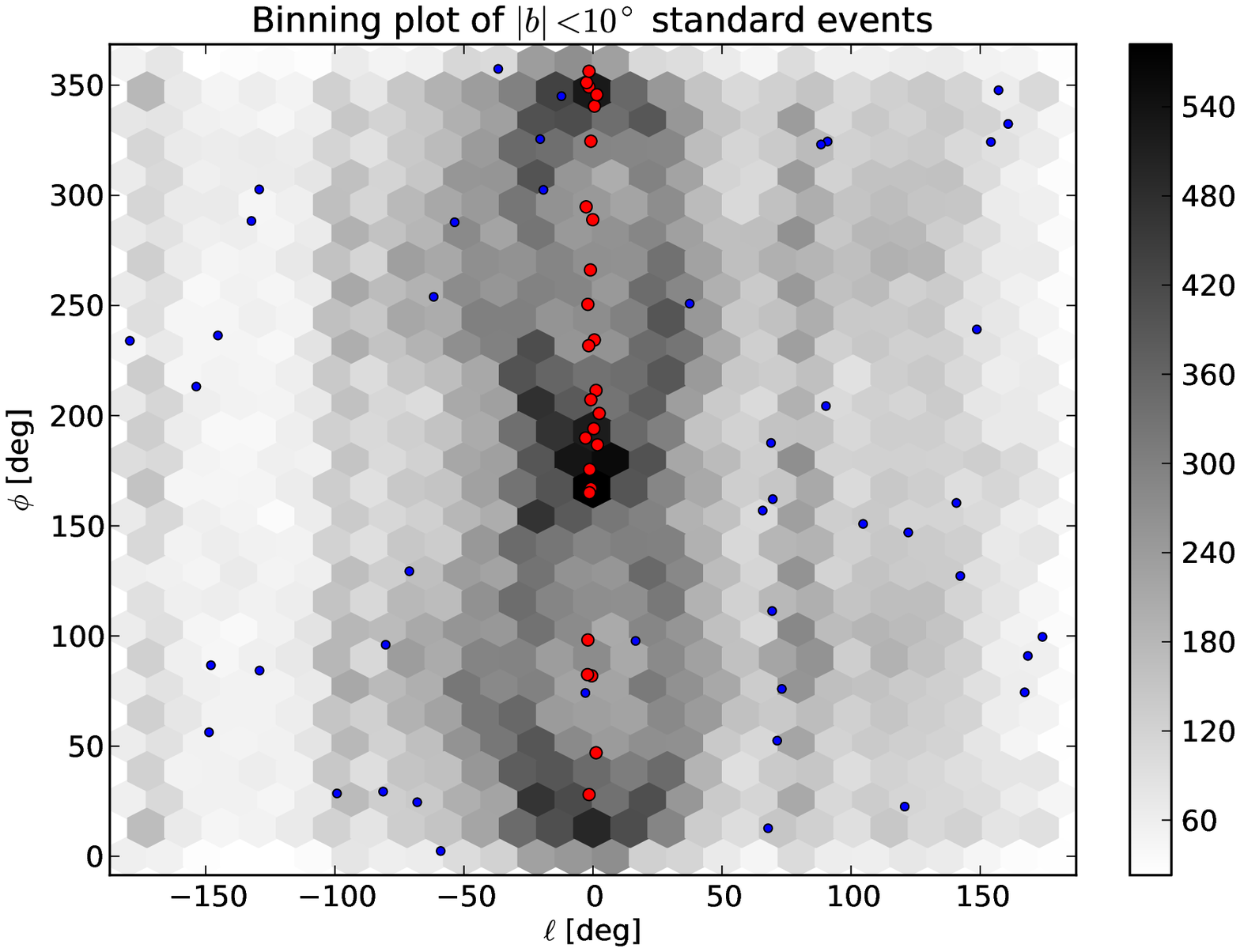}
  \includegraphics[width=0.44\textwidth]{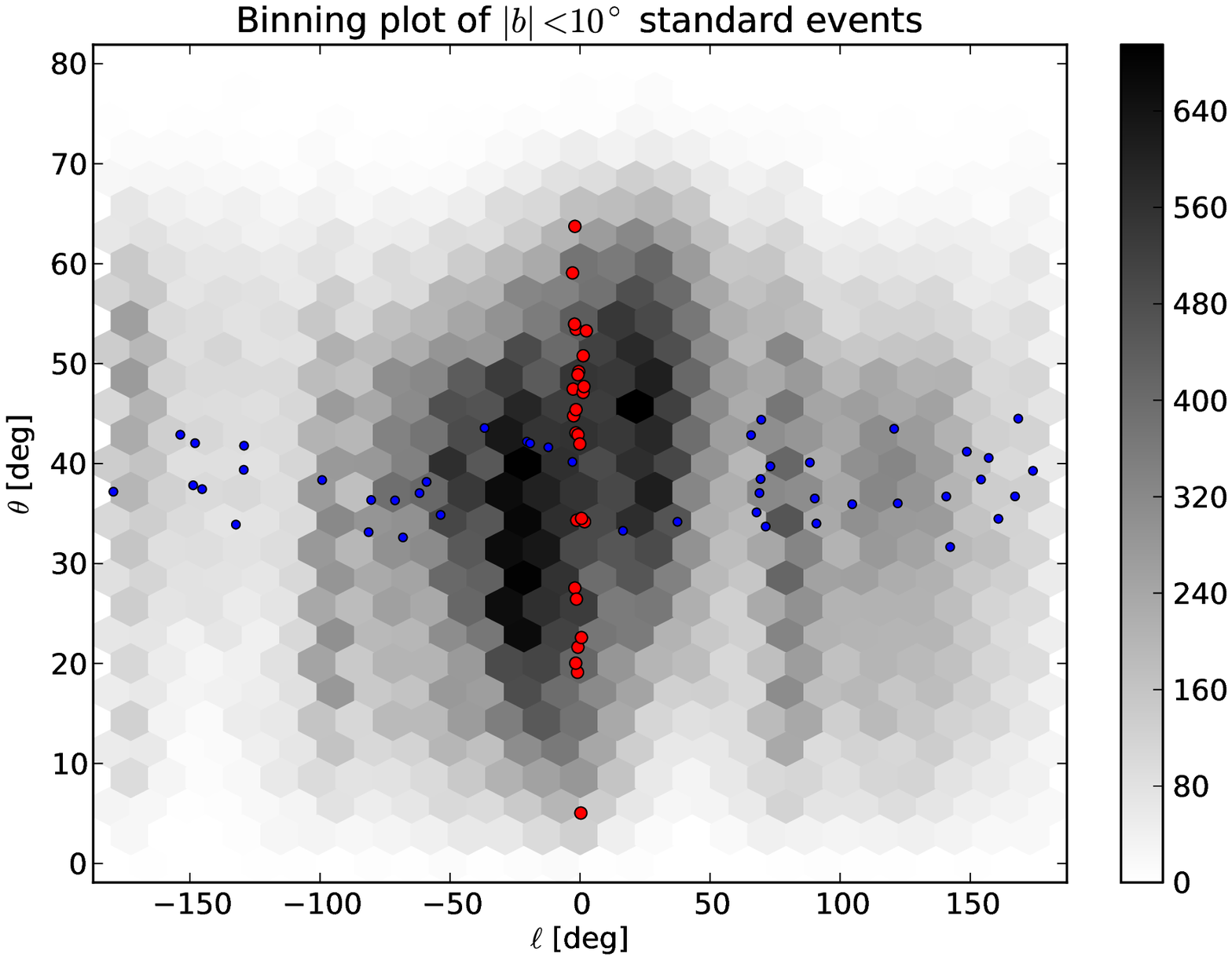}
  \caption{\emph{Upper panels:} $\phi$ (left panel) and
  $\theta$ (right panel) distributions of standard events
  close to the Galactic center, as a function of time modulo
  one year (with Jan 1st at the origin). In Dec (Jun) the
  Sun passes near the Galactic center (anti-center).  In
  these periods, most GC events are observed at $\phi\approx
  0^\circ$ ($\phi\approx 180^\circ$), since the LAT $x-z$
  plane is determined by the Sun to keep the solar panels
  oriented. \emph{Lower panels:} $\phi$ (left panel) and
  $\theta$ (right panel) distribution of standard events
  along the Galactic disk. Close to the GC, the distribution
  becomes significantly bimodal. The red and blue dots are
  as in Fig.~\ref{fig:phiThetaDist}.}
  \label{fig:time_phi}
\end{figure*}

\begin{figure}
  \centering
  \includegraphics[width=1.0\linewidth]{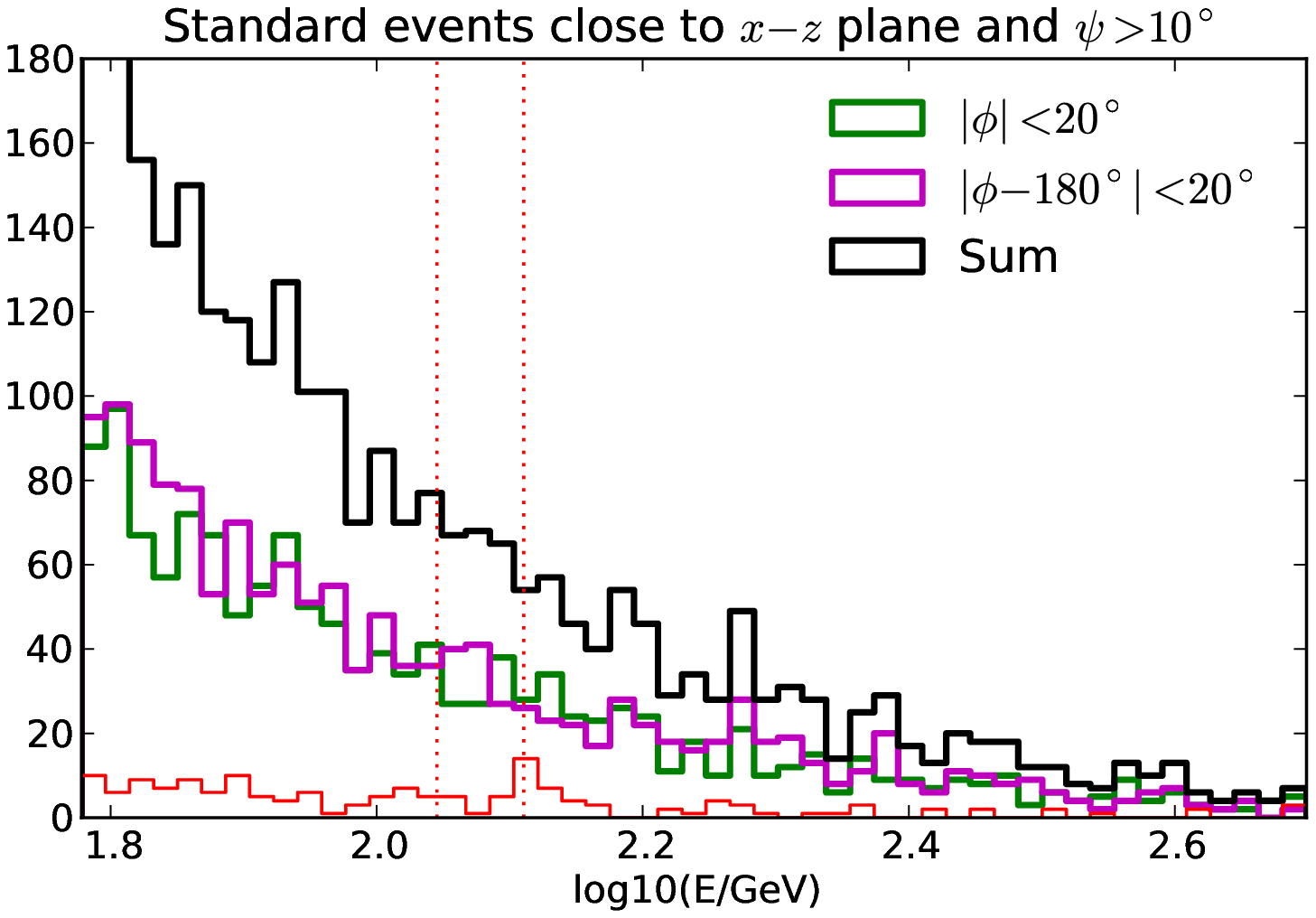}
  \caption{Energy distribution of standard events away from the GC with
  incidence angles close to the LAT $x-z$ plane (perpendicular to the solar
  panels). The red dotted lines indicate 111 and 129 GeV; the thin red line
  shows GC events for comparison.}
  \label{fig:spectrum_phi}
\end{figure}

\subsubsection{Hypothesis: The Galactic
center has a hard spectrum, making energy mapping errors more
significant.}

The Galactic center black hole (Sgr A$^*$) is visible up to 20 
TeV, and other
sources in the GC may be unusually hard. If these high energy
photons are occasionally mis-reconstructed with energies
close to 130 GeV, this could produce a line feature in the
data that would appear preferentially at the Galactic center.
In Table~\ref{tab:regions}, we list the ratio of $E>300\GeV$
events to $E>100\GeV$ events and of $E>100\GeV$ to $E>30\GeV$
events for our samples.  The
GC spectrum is not much harder than the rest of the Inner Galactic
plane, but the latter shows no sign of a feature at 130 GeV
(Fig.~\ref{fig:spectra1}). Note that
about $\sim20$ events contribute to the central part of the
line feature seen at the GC~\citep{linepaper}. 
Assuming that the GC spectrum is not
harder than $dN/dE \propto E^{-2.0}$ (normalized to the
number of events above 150 GeV), the GC events cannot come
from energies above 300 GeV, since there would not be enough
events to make up the 130 GeV excess even if \emph{all} of
them were incorrectly mapped to 130 GeV. The possibility of energy
mapping errors of events close to 130 GeV will be discussed
in Sec.~\ref{sec:EarthLimb}.

\subsubsection{Hypothesis: the GC observations have a
restricted range of incidence angles on the instrument.}

If the Galactic center were predominantly observed at
specific angles in LAT instrumental coordinates, associated
instrumental problems could be projected onto the Galactic
center simply for geometric reasons. 
In late December (June), the Sun crosses the Galactic disk,
and the angular
distance between the Sun and the Galactic center
(anti-center) decreases to $\sim5^\circ$. Since \Fermi\ keeps
the solar panels aligned to the Sun (the Sun is in the $x-z$ plane), this
leads to an
increase of Galactic center events at instrument azimuth
angles of $\phi\approx 0^\circ$ ($\phi\approx 180^\circ$).
This behaviour is clearly visible in the upper left panel of
Fig.~\ref{fig:time_phi}, where we show the $\phi$
distribution of $>10\GeV$ events from the Galactic center as
a function of time of year.  The $\theta$
distribution in the upper right panel reflects the
precession pattern with a $\sim53.4$ day period. Integrated
over time, the $\phi$ and $\theta$ distributions look like
in the lower panels of Fig.~\ref{fig:time_phi}, where they
are shown as a function of the Galactic longitude. Close to
the GC, the $\phi$ distribution becomes significantly
bimodal. Note that this effect occurs along the full 
ecliptic,
but the intersection with the Galactic disk
close to the center is accompanied by the largest number of
events.

It is tempting to relate the location of the 130 GeV excess
in the Galactic plane to this inhomogeneous $\phi$
distribution. As a test, we select standard events from the
full sky (excluding $\psi < 10^\circ$) in the range $\phi=
-20\dots20^\circ\ \text{mod}\ 180^\circ$. The total number
of these events is $\sim10$ times larger than the number of
Galactic center events; an anomaly in the event
reconstruction near these $\phi$ angles would appear in
this data sample with high significance.  However, the
energy distribution shown in Fig.~\ref{fig:spectrum_phi}
shows no significant feature at 111 or 129 GeV.

\begin{figure*}
  \centering
  \includegraphics[width=0.45\textwidth]{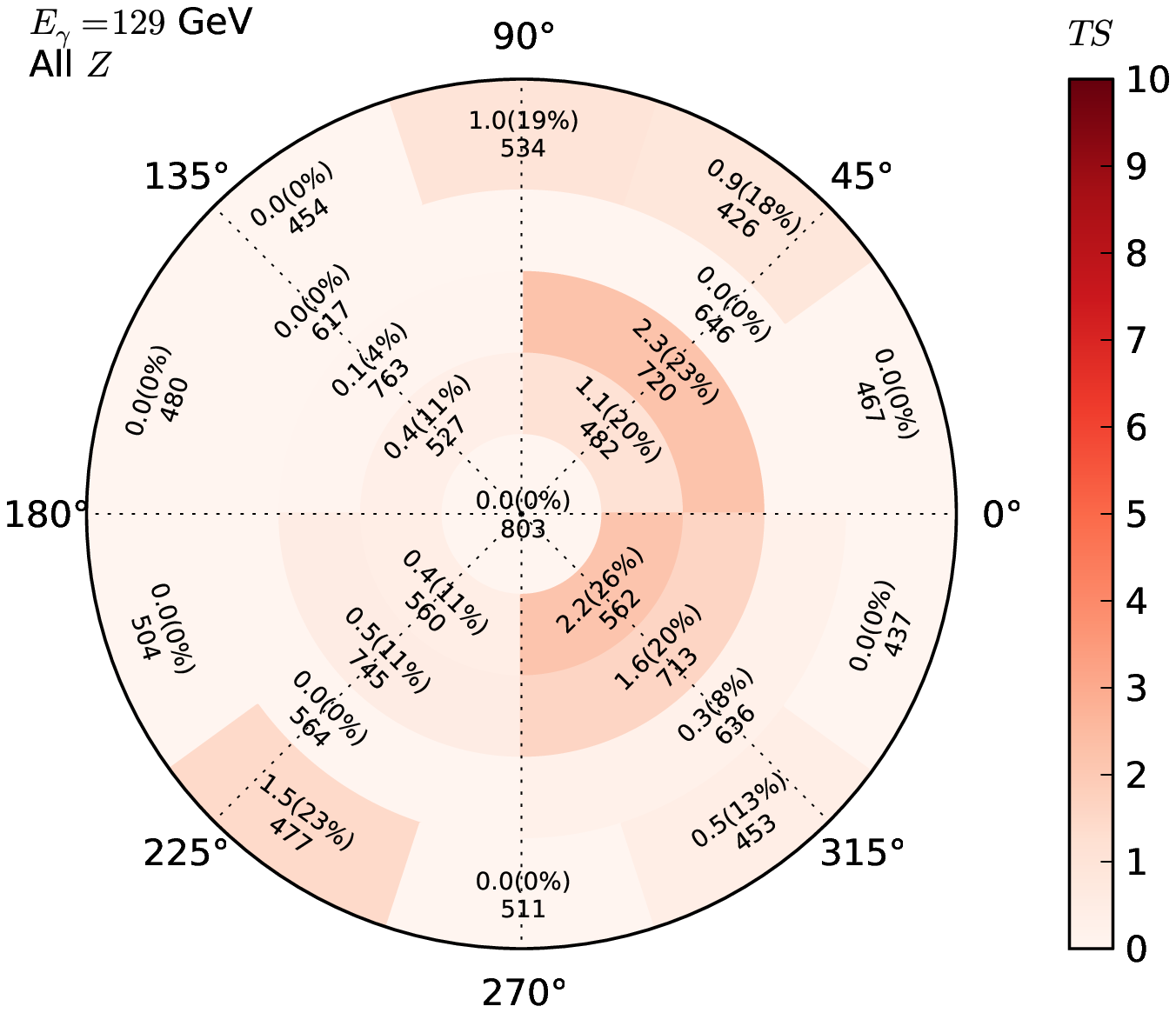}
  \includegraphics[width=0.40\textwidth]{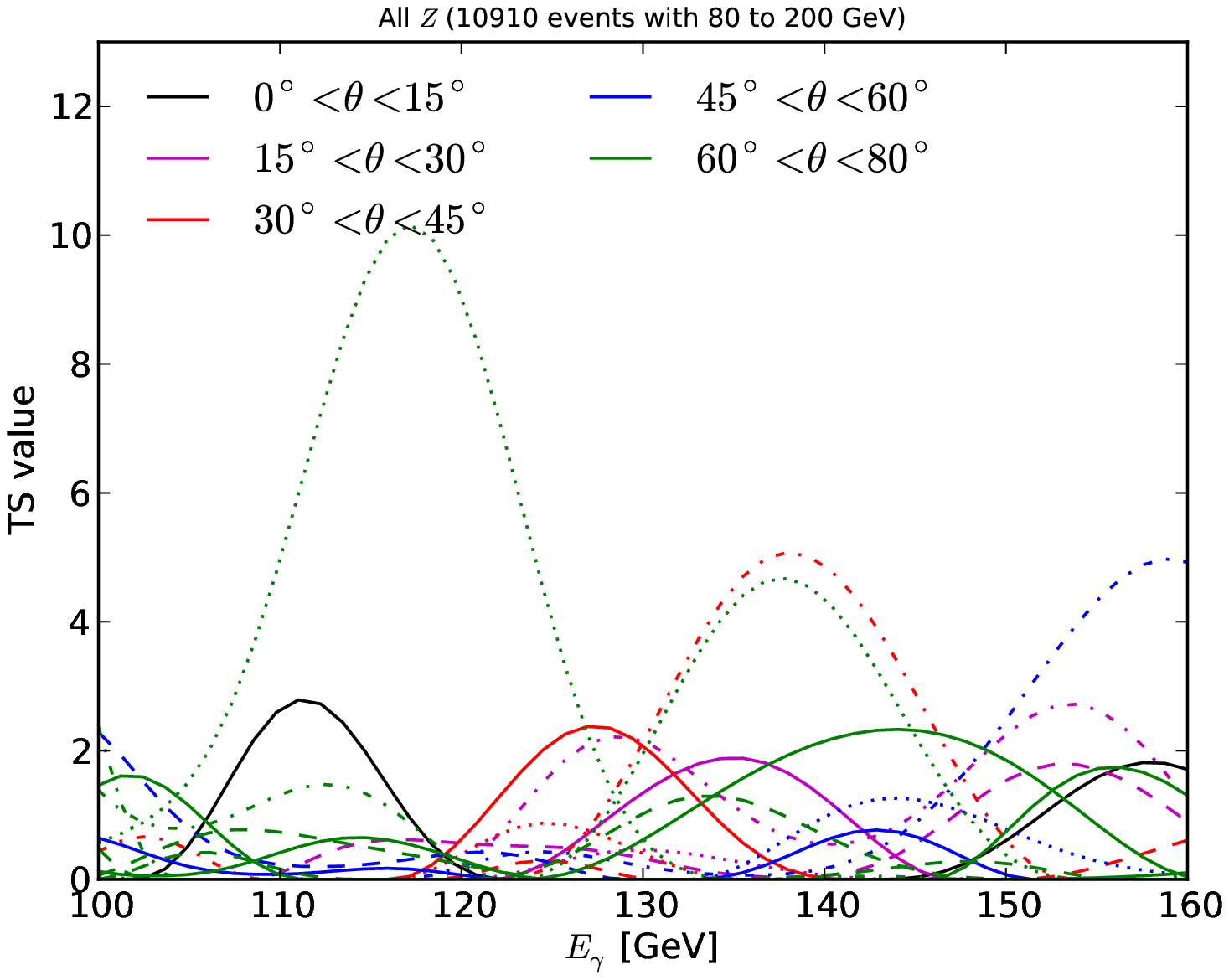}
  \includegraphics[width=0.45\textwidth]{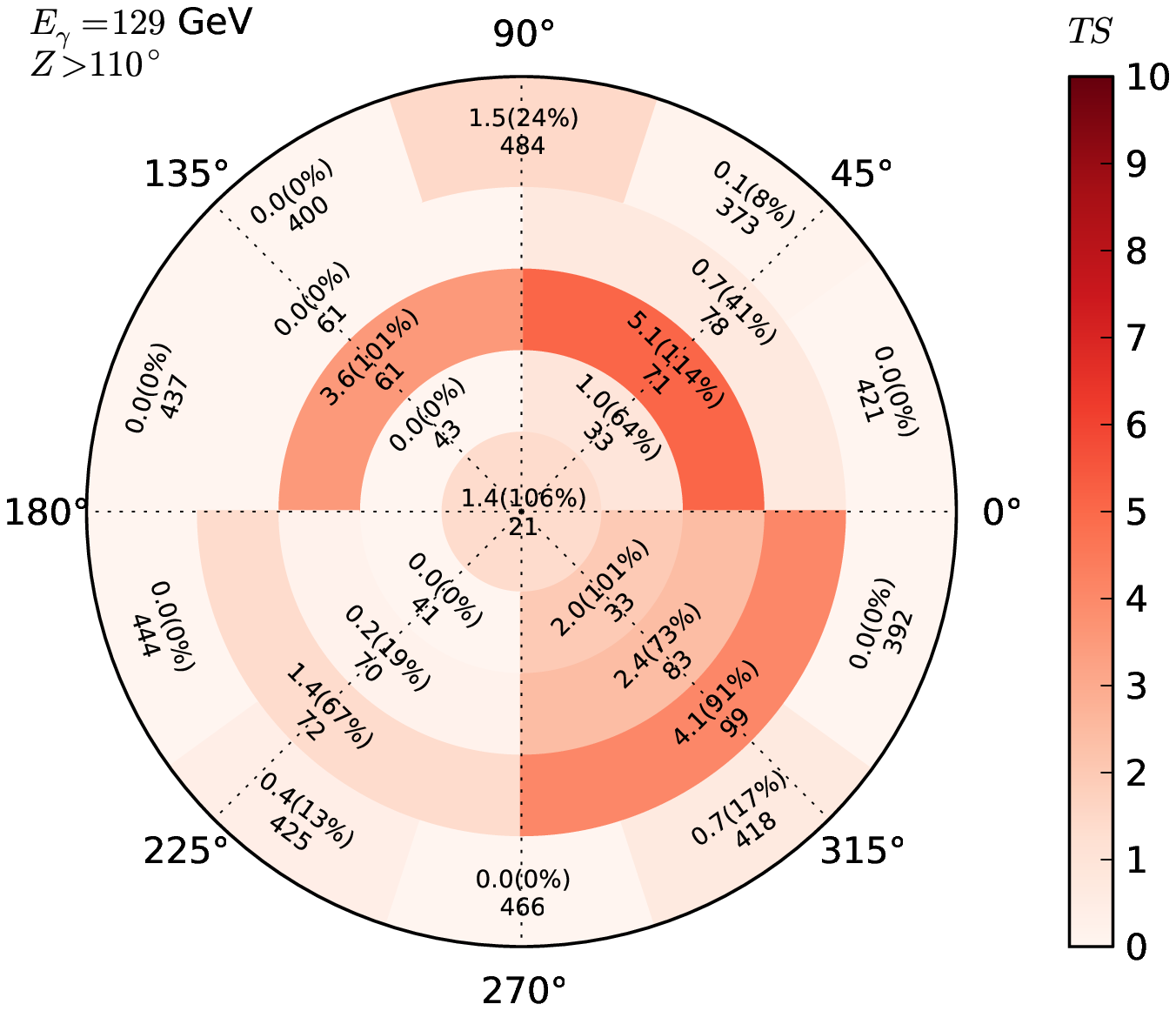}
  \includegraphics[width=0.40\textwidth]{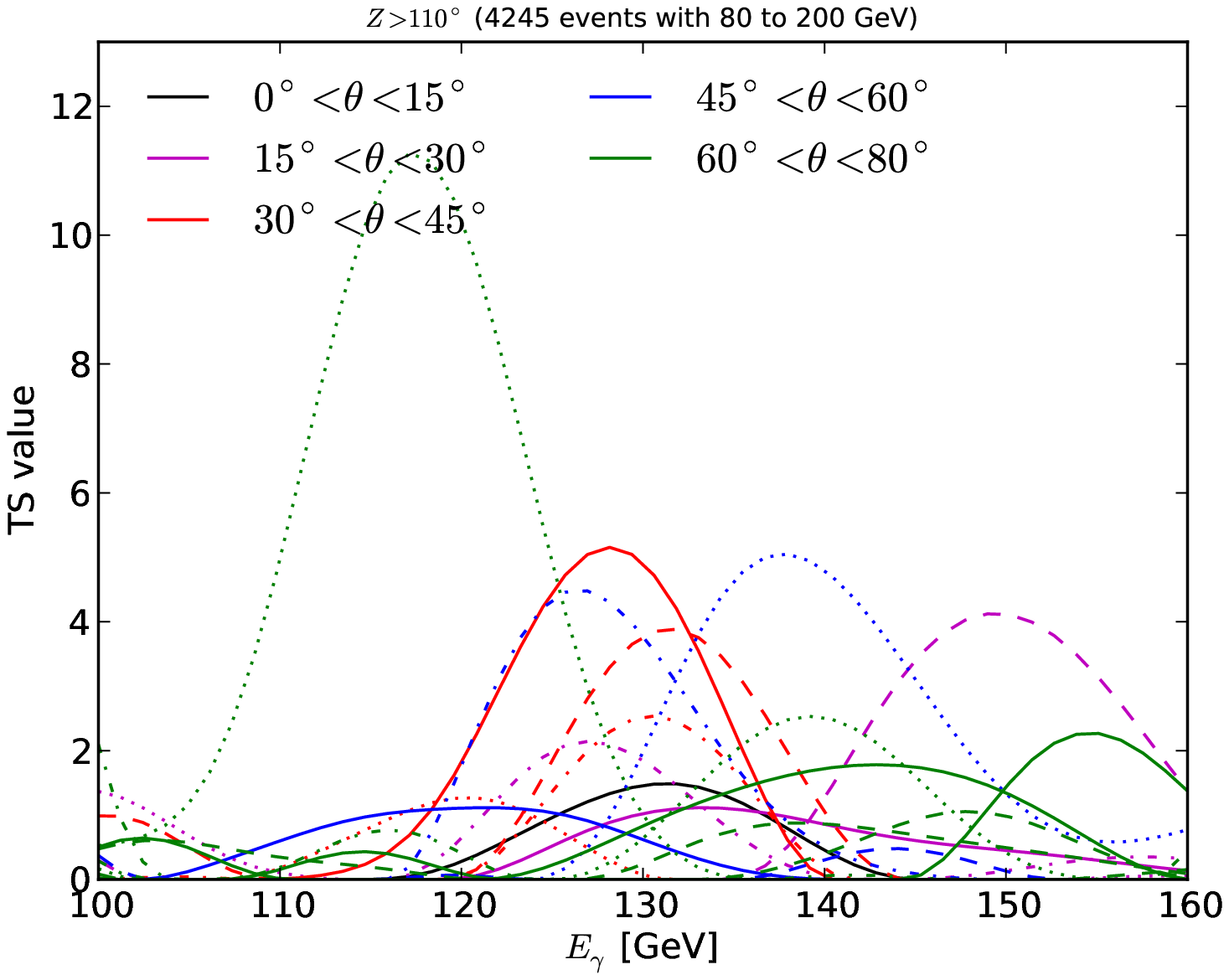}
  \includegraphics[width=0.45\textwidth]{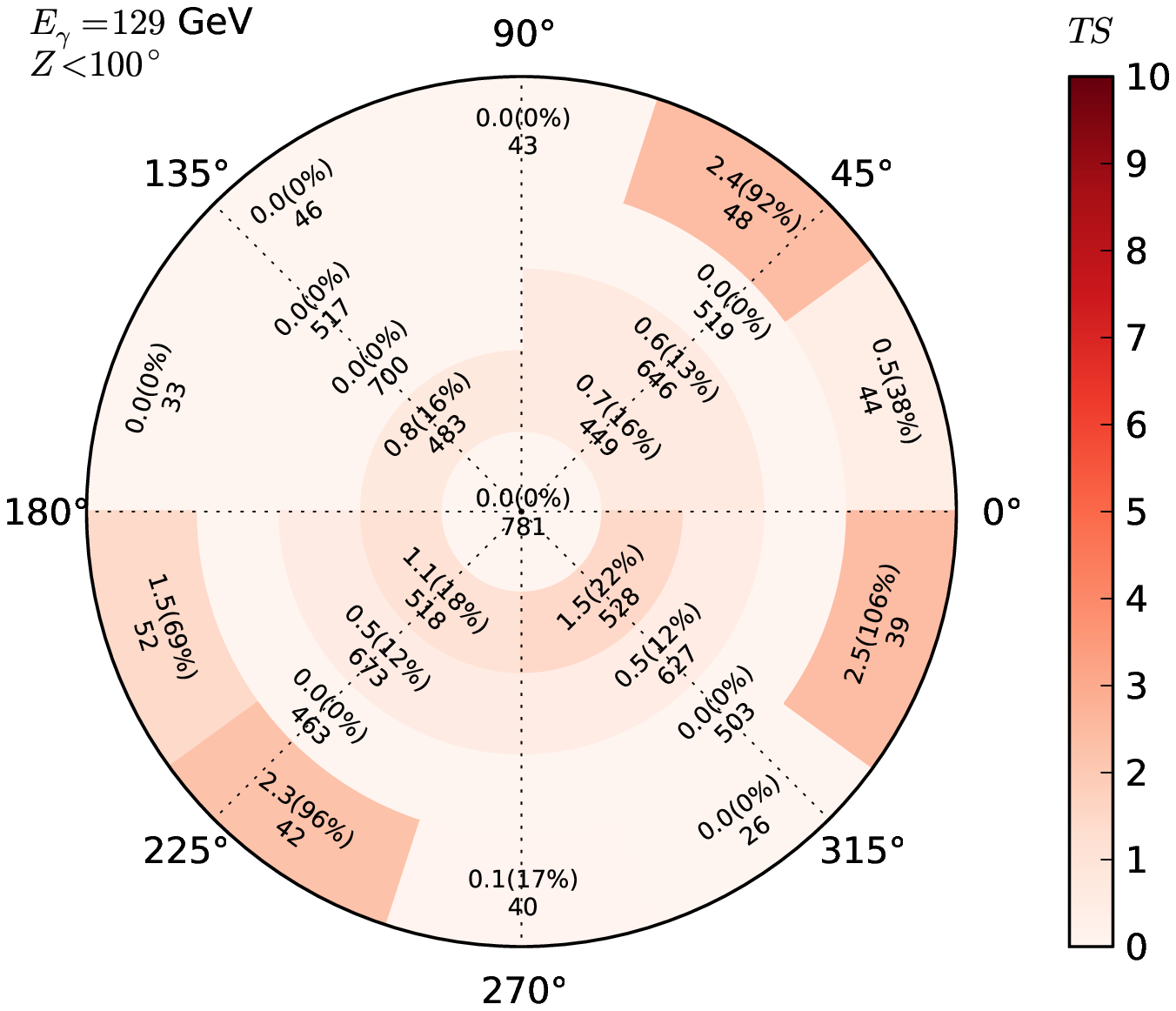}
  \includegraphics[width=0.40\textwidth]{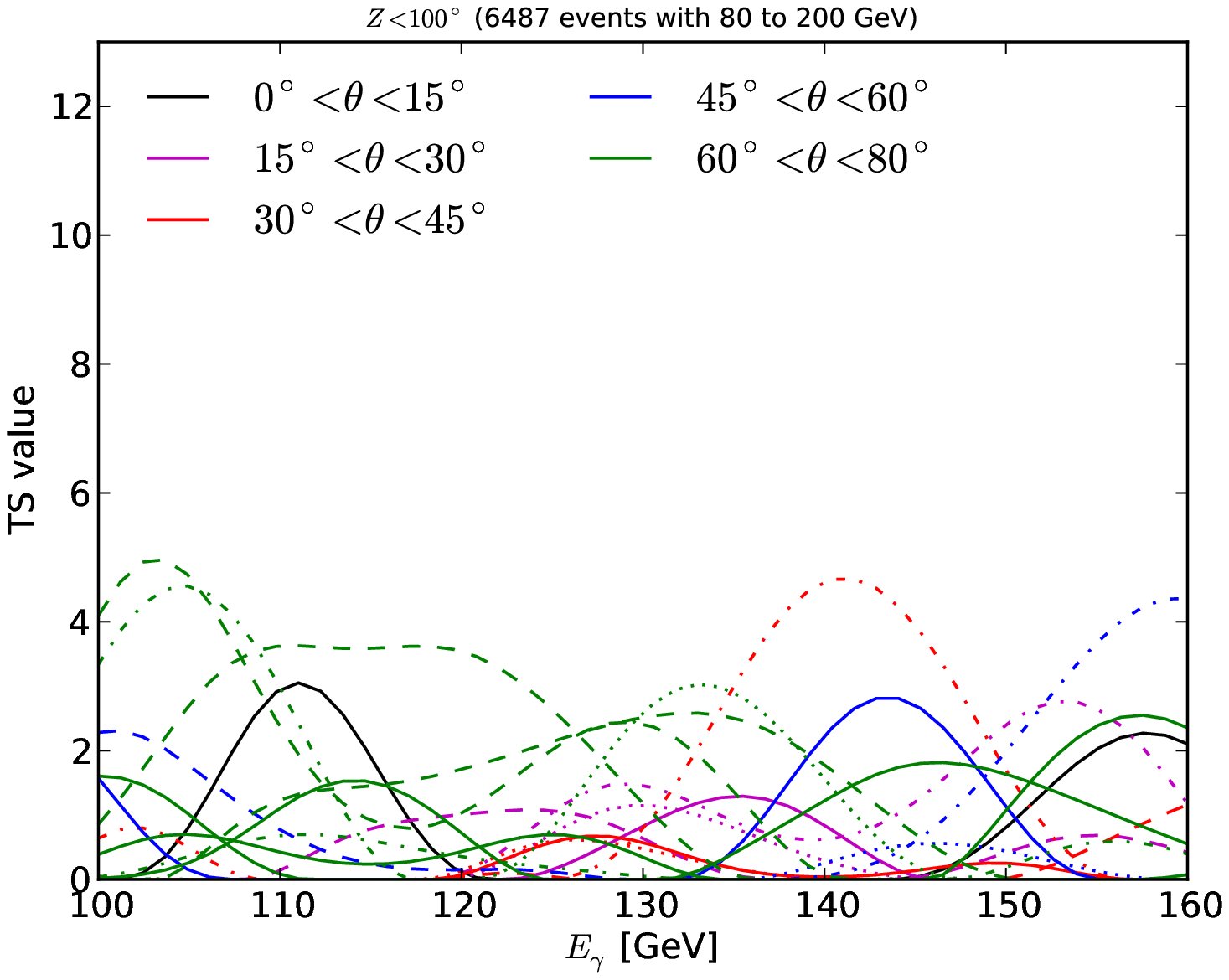}
  \caption{\emph{Left panels:} Significance for a $\sim130$ GeV
  line-like excess in different parts of the $\theta$-$\phi$
  plane (for $\theta=0$--$80^\circ$). From top to bottom,
  the results are derived from all events, from the Earth
  limb events, and from standard events. The three numbers
  show the TS value for the presence of a 129 GeV line, the
  signal-to-background ratio and the number of events above
  $80\GeV$ inside the considered $\theta$-$\phi$ region.
  $\theta$-cuts are made at $15^\circ$, $30^\circ$, $45^\circ$ and $60^\circ$.
  \emph{Right panels:} The significance for a
  line-like excess as a function of $E_\gamma$ for the same $\theta$-$\phi$ regions.
  Solid (dashed, dotted, dash-dotted) lines correspond to panels at different
  $\phi$, starting counterclockwise from $\phi=0^\circ$.
  }
  \label{fig:polarPlotsAll}
\end{figure*}

\subsection{Peculiarities of the instrument}
Instrumental effects are likely to be correlated with specific instrumental
coordinates rather than sky coordinates. In this subsection we search for
suspicious trends at 130 GeV as a function of the event incidence angles,
quality parameters, and arrival times.  We also search for the possibility of
`hotspots' at other regions of the sky; they could indicate instrumental
effects or an exotic source population.
\medskip

An error in the LAT effective area (e.g.~various cuts at $\sim130\GeV$ are
less/more restrictive for some reason) 
could
in principle explain the GC line, but would also produce a line in the other
samples listed in Tab.~\ref{tab:regions}. 
Exactly this already happened in the past at energies close to
$10\GeV$~\cite{Ackermann:2012qk}.
However, no 130 GeV feature
is seen in the main test samples outside of the GC (though in a small subsample
that we will discuss below). The LAT team
estimated the point-to-point correlation of the effective area on scales
relevant for line searches to be on the level of
$2\%$~\cite{Fermi:caveatsPASS7},
which is not enough to explain the $\mathcal{O}(1)$ GC or Earth limb features.

If the GC line is due to non-rejected CRs, it is hard to understand how
mono-energetic particles could be present in the CR spectrum. 
What is more,
a bright region like the Galactic center has a large gamma-ray-to-CR
ratio, such that a CR contamination would affect it last, not first.

\subsubsection{Hypothesis: A 130 GeV features is visible
at specific incidence angles on the instrument.}


One concern is that a 130 GeV feature in the
reconstructed events is visible only for events with certain
incidence angles $(\theta, \phi$) in instrumental
coordinates. The 130 GeV excess would be then imprinted on
regions of the sky that are predominantly observed at
these problematic angles.\footnote{A study of the
$x$--$y$ positions of the events inside the tracker and calorimeter
is not possible using public data only.
It is however difficult to conceive how these 
parameters could be correlated with certain regions of the
sky.}

To study line-like features at different incidence
angles, we split up the $(\theta, \phi)$ plane in different
regions as shown in the left panels of
Fig.~\ref{fig:polarPlotsAll}. We then analyse the spectrum
of all events that hit the detector at these incidence
angle patches separately and search for lines. The fits are
performed as in Ref.~\cite{Weniger:2012}, but the energy
window is fixed to 80 to 200 GeV, we assume a flat
acceptance and for a line we take a simple Gaussian with a 
standard deviation of $6\%$ (to approximate the central part of the non-Gaussian
LAT energy dispersion).

The left panels of Fig.~\ref{fig:polarPlotsAll} show from
top to bottom the results obtained (1) using all events, (2)
using Earth limb events only, and (3) using standard events
only; the latter two are disjoint subsets of the
former one. For each $(\theta, \phi)$ range we quote three
numbers: the significance of a 129 GeV line-feature in terms
of the $TS$ value, the signal-to-background ratio of the
putative line signal, and the number of events above 80 GeV
that contribute to this region. The $TS$ values are also
indicated by color for better visibility.


We also compute the $TS$ as a function of energy in each region of
$(\theta,\phi)$ to check whether there is anything unusual about 130 GeV
(Fig.~\ref{fig:polarPlotsAll}, right panels). In
particular the lower two plots which correspond to
independent data sets show clearly that the enhanced $TS$
values at different energies in different panels are not
correlated. Lastly, we note that the significance of the
largest $TS$ value, $TS\simeq11$ (at $115\GeV$), has a $p$-value of
about $\sim0.3$ when taking into account
$\sim2\times23\times2$ trials over a $\chi_{k=2}^2$ distribution
(two $Z$ ranges, $23$ incidence angle panels, $\sim2$
independent search regions~\cite{Gross:2010qma}). 

Fig.~\ref{fig:polarPlotsAll} suggests that the TS values observed in some panels in case of
Earth limb ($Z>110^\circ$) events are well within the statistical expectations
once the large number of associated trials is taken into account.
However, it cannot be excluded that the 130 GeV excess in
these parts of the Earth limb indicates a real instrumental
effect. Then the energy
reconstruction at different incidence angles would have to somehow
additionally
depend on the zenith angle $Z$ of the events
(e.g.~via a
subtle correlation with the rocking angle).
We postpone an in-depth discussion of this possibility to
section~\ref{sec:EarthLimb}; in any case more
Earth limb data will help to verify or falsify this possibility.

\subsubsection{Hypothesis: The Galactic center events are flagged as badly
reconstructed.}

\begin{figure}
  \centering
  \includegraphics[width=1.00\linewidth]{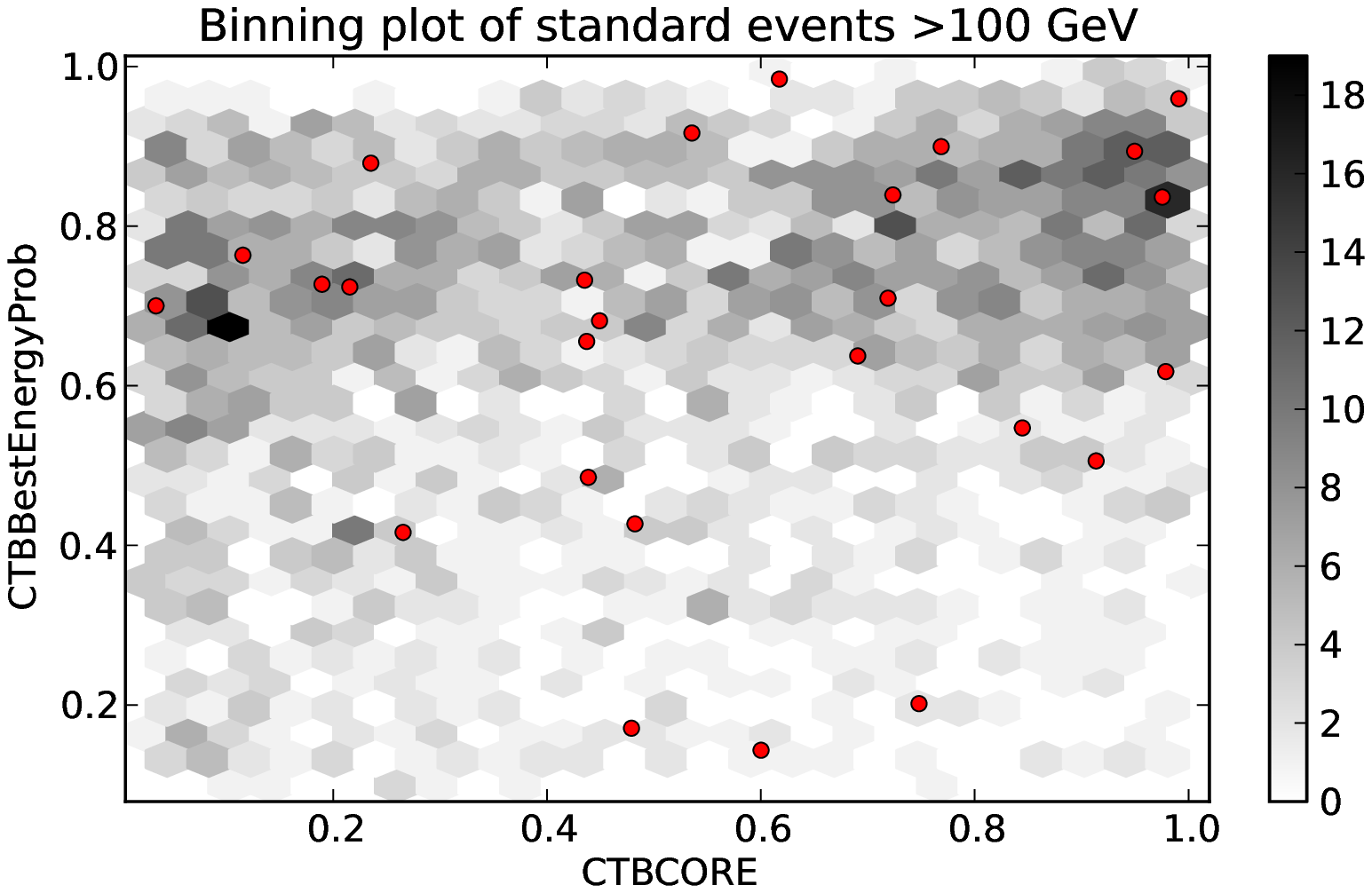}
  \includegraphics[width=1.00\linewidth]{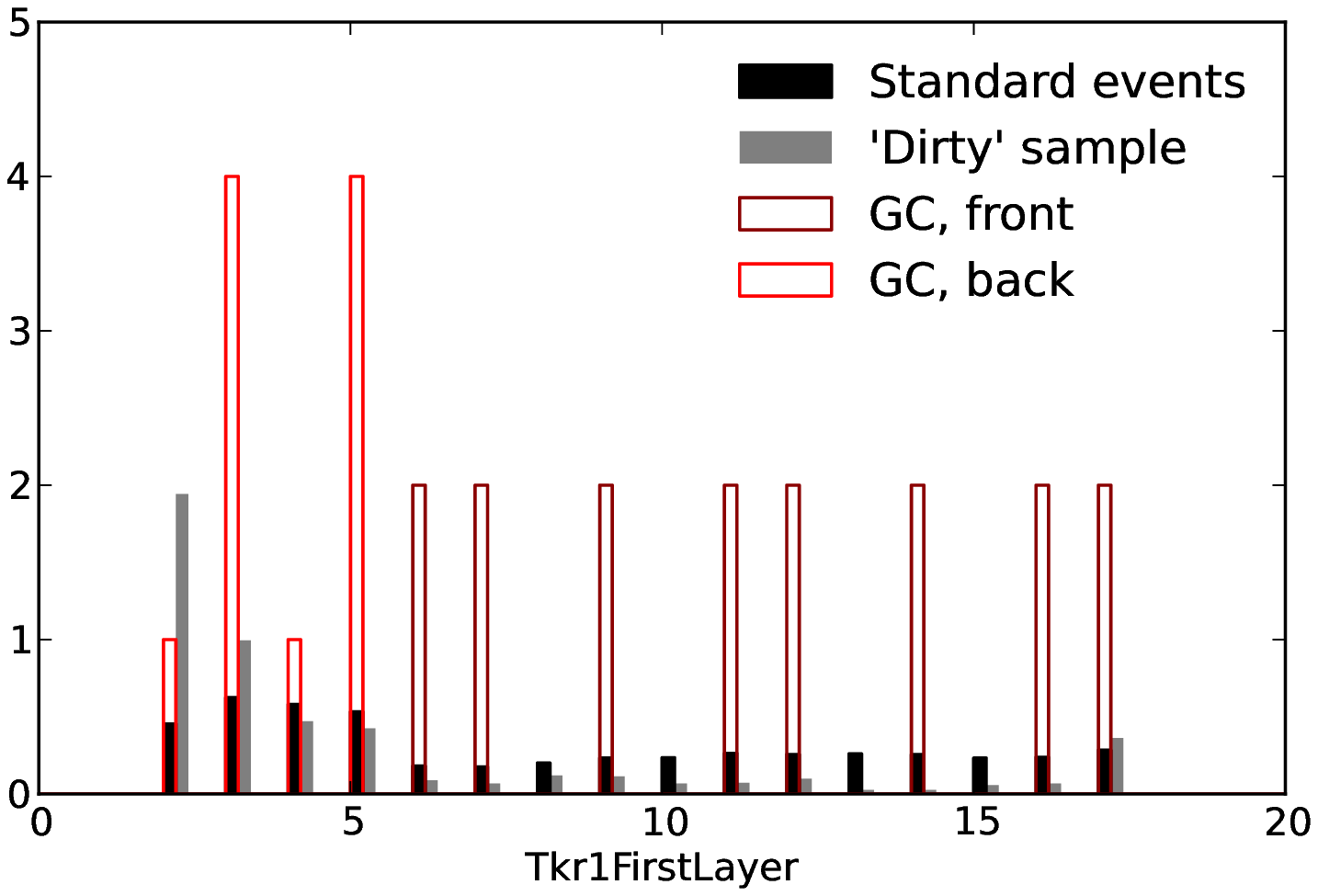}
  \caption{\emph{Upper panel:} Distribution of CTBCORE (the
  probability that the direction estimate is good) and
  CTBBestEnergyProb (the probability that the best energy
  chosen from the two energy estimators is correct) for
  Galactic center line events (in red), compared to the
  distribution of $>100$ GeV standard events. \emph{Lower
  panel}: First tracker layer to show evidence of a particle
  hit for the best track reconstruction. Tracker layers are
  0-17 where 0 is closest to the calorimeter, and 6-17 (2-5)
  corresponds to \texttt{FRONT}- (\texttt{BACK}-)converting events (tracker
  layers 0 and 1 have no conversion foils). The dark (light) gray
  bars show the average over all $>100$ GeV
  standard events ('dirty' sample defined as \texttt{SOURCE-CLEAN} events with
  $|b|>5^\circ$); the red bars show the distribution for the GC line.}
  \label{fig:CTBquality}
\end{figure}

Although we do not have access to all details of the event reconstruction, the
extended LAT event files contain a few figure-of-merit quantities from the first step of the
event-level analysis, which inform about the quality of
calorimeter and tracker event reconstruction.\footnote{\url{http://fermi.gsfc.nasa.gov/ssc/%
data/analysis/documentation/Cicerone/Cicerone\_Data/LAT\_Data\_Columns.html}}
CTBCORE describes the probability that the direction estimate
is good (roughly the probability that the reconstructed
direction falls within the nominal 68\% containment angle),
CTBBestEnergyProb the probability that the reconstructed
energy falls into the core of the energy
dispersion~\cite{collaboration:2012kca}. We show
these parameters for the the GC line events in the upper
panel of Fig.~\ref{fig:CTBquality}; the background histogram
shows the distribution of the these parameters in standard
events above 100 GeV. No significant bias in the distribution
appears.

The red bars in the lower panel of
Fig.~\ref{fig:CTBquality} indicate the first tracker layer that
shows evidence of a particle hit for the best track
reconstruction (Tkr1FirstLayer) in case of the GC line
events.  Tracker layers are 0--17, where 0 is closest to the
calorimeter and 6--17 (2--5) corresponds to \texttt{FRONT}-
(\texttt{BACK}-)converting events (tracker layers 0 and 1 do not have
conversion foils due to requirements of the three-in-a-row trigger
primitive). The dark gray bars show the
distribution averaged over all $>100$ GeV standard events.
The light gray bars
show for comparison the distribution of a `dirty' $>100$~GeV event sample.
It is defined as all \texttt{SOURCE-CLEAN} events with $|b|>5^\circ$,
and hence has a high
CR contamination. The distribution of the latter is significantly biased
towards the first and last tracker layers, whereas
the distribution of GC line events is compatible with the
expectations for standard events.

\subsubsection{Hypothesis: There are `hotspots' with
line-like features in other sky regions.}

\begin{figure}
  \begin{center}
    \includegraphics[width=1.0\linewidth]{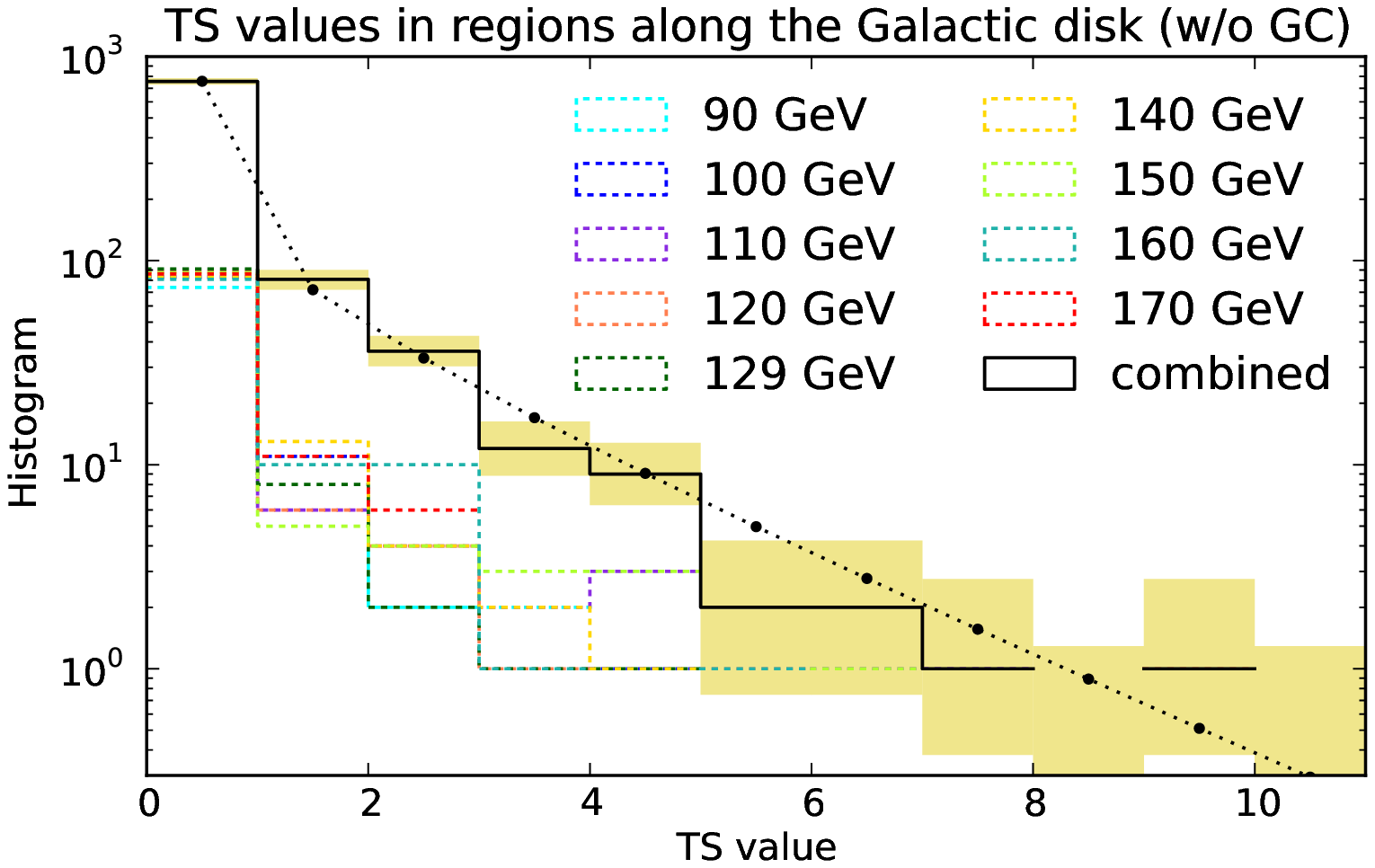}
  \end{center}
  \caption{Histogram of TS values observed in partially overlapping
  $6^\circ\times6^\circ$ regions along the Galactic disk, centered on $b=0^\circ$ and
  $|\ell|=13.5^\circ, 16.5^\circ \dots 178.5^\circ$. We show results for
  different line energies from 90 to $170\GeV$ (dotted lines), as well as all
  TS values combined (black solid line); in the latter case, the black dotted
  line is the theoretical expectation (a $0.5\chi^2_{k=0}+0.5\chi^2_{k=1}$
  distribution), the yellow band shows the $\pm1\sigma$ errors. The tail of the
  observed distribution looks as expected.}
  \label{fig:hotspots}
\end{figure}

The 130 GeV excess is located in a narrow range of about $\sim5^\circ$ around
the Galactic center. The existence of other significant `hotspots' (localized
regions with a significant line-like emission at or around 130 GeV) in the sky
could indicate an unexpected instrumental effect, or point to a new exotic
source class~\cite{Boyarsky:2012ca}. 
In absence of knowledge about the distribution of such possible
sources, we adopt the following strategy. We select partially overlapping regions of
$6^\circ\times6^\circ$ size, centered along the Galactic disk at
$b=0^\circ$, $|\ell|=13.5^\circ, 16.5^\circ \dots 178.5^\circ$. The size of
the region is chosen to include roughly 50--100 events above 80 GeV (like in our `GC
region'); regions close to the GC are excluded to avoid contamination with the
130 GeV excess. In each region we calculate the TS value for the presence of
a line at various energies. The details of the fit are the same as above for
Fig.~\ref{fig:spectra1} (in particular the energy window is always fixed to
80--210~GeV). The distribution of TS values obtained in this way is
shown in Fig.~\ref{fig:hotspots} by the dotted colored lines for different
line energies; a combination of
TS values for all line energies is shown by the solid black line. Its tail
distribution is in excellent agreement with the statistical expectations
(shown by the black dotted line) within the $\pm1\sigma$ error bars, showing
no indication for the presence of `hotspots'. We obtain the same 
result when reducing the size of the regions,
shifting them by a common offset, or selecting random circular regions all
over the
sky with the requirement that $\sim100$ events are included. 

\subsubsection{Hypothesis: The observed signal is variable.}

\begin{figure}
  \begin{center}
    \includegraphics[width=1.0\linewidth]{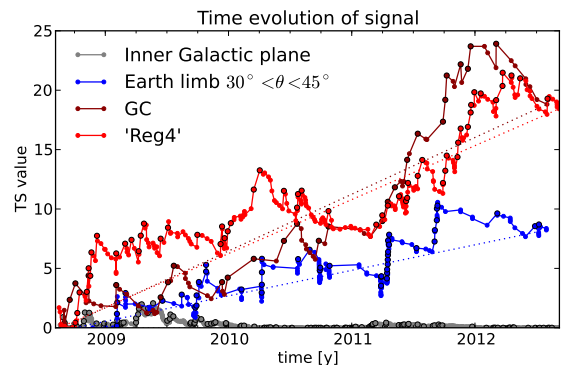}
  \end{center}
  \caption{Time evolution of TS values. In dark (light) red we show results
  for the GC region from Tab.~\ref{tab:regions} (respectively Reg4 from
  Ref.~\cite{Weniger:2012}), in blue we show the evolution of the Earth limb
  line. The gray line indicates for comparison the TS value obtained for the
  Inner Galactic plane, where no signal is observed.  Jumps in the Earth limb
  line significance are related to times when
  \zrock$>52^\circ$, cp.~Fig.~\ref{fig:timehist}.}
  \label{fig:timeevolution}
\end{figure}

An interpretation of the 130 GeV excess in
terms of dark matter annihilation requires steadiness of the source;
a strong variability could
indicate (time-dependent) instrumental effects. In
Fig.~\ref{fig:timeevolution} we plot how the $TS$ value of
the GC signal evolved over time (assuming $E_\gamma=129.8\GeV$~\cite{Weniger:2012}). 
The dark red line corresponds
to the GC region from Tab.~\ref{tab:regions}, the light red line shows the
results for region Reg4 from Ref.~\cite{Weniger:2012}. We
compare them to the time-evolution of the suspicious Earth limb
line (blue) and the time evolution of the TS value obtained from the
Inner Galactic plane (gray). The `GC region' curve appears to have grown most
strongly between March 2011 and February 2012, and falling
during last few months; on the other hand, the signal observed in the larger Reg4
does not show the same behaviour and appears more steady. In all cases, the open circles
indicate an event between 120 and 138 GeV.
The curves show no strong sign for a variability of the line features.  Note
that the number of GC region line events in the first and second half of the
observational period are respectively 12 and 14.

%

\section{Earth limb photons}
\label{sec:EarthLimb}

\begin{figure}
  \centering
  \includegraphics[width=1.0\linewidth]{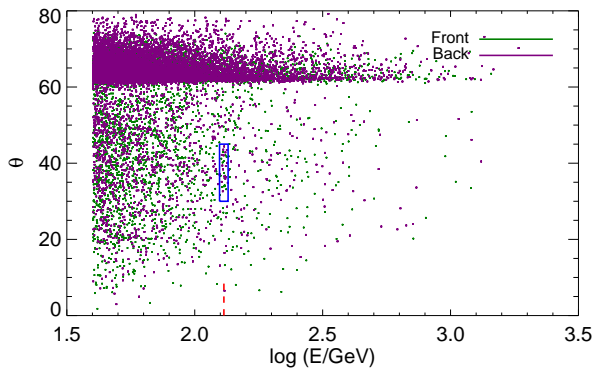}
  \caption{Incidence angle $\theta$ vs. log $E$ for $Z >
    110\degree$, with \texttt{FRONT} (dark green) and \texttt{BACK} (purple)
    events. A blue box indicates the region $30\degree <
    \theta < 45\degree$ and 125 GeV $< E <$ 135 GeV, where
    an excess of events appear.  The vast majority of limb
    events are at $\theta > 60\degree$ because the telescope
    seldom points more than $50\degree$ from zenith, and the
    limb events are mostly at $Z>110\degree$. }
  \label{fig:theta-E-frontback}
\end{figure}

\begin{figure}
  \centering
  \includegraphics[width=1.0\linewidth]{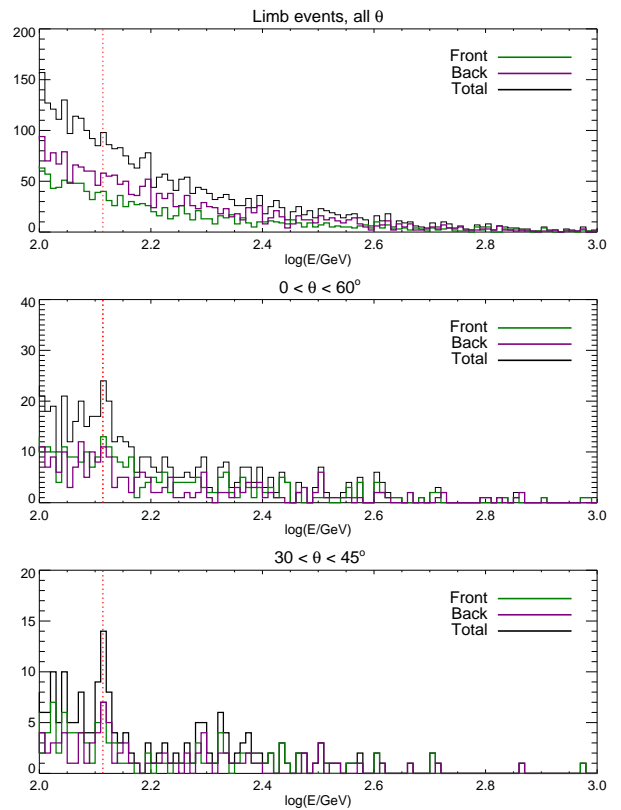}
  \caption{Histogram of Earth limb $(Z>110\degree)$ events vs.
    log($E$) for various ranges of $\theta$. 
    $E=129\GeV$ is
    indicated by red dotted line.  Note the excess in the
    $30\degree < \theta < 45\degree$ incidence angle bin,
    which contains about 6\% of the limb
    events. \texttt{FRONT} and \texttt{BACK} converting
    events are overplotted with dark green and purple lines
    respectively. The 130 GeV excess appears equally in
    \texttt{FRONT} and \texttt{BACK} converting events.}
  \label{fig:Ehist-all}
\end{figure}

\begin{figure}
  \centering
  \includegraphics[width=1.0\linewidth]{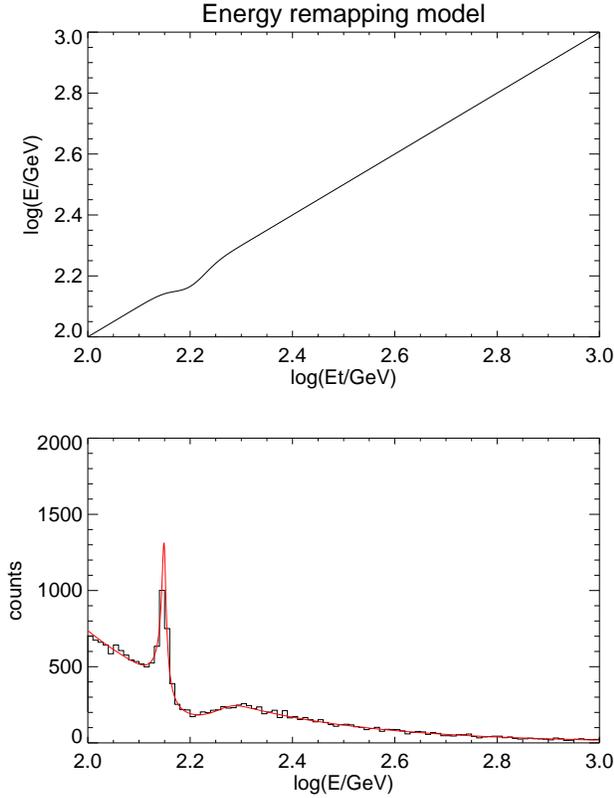}
  \caption{Upper panel: function mapping true energy $E_t$
  to reported energy $E$ (see Eqs. \ref{eq:yofx} and
  \ref{eq:dydx}). Lower panel: The effect of this mapping
  on a spectrum $dN/dE \sim E^{-2.6}$,
  as in Eq.
  \ref{eq:dndy} (red line) and also for mock data (black
  histogram).}
  \label{fig:bumpmodel}
\end{figure}

\begin{figure}
  \centering
  \includegraphics[width=1.0\linewidth]{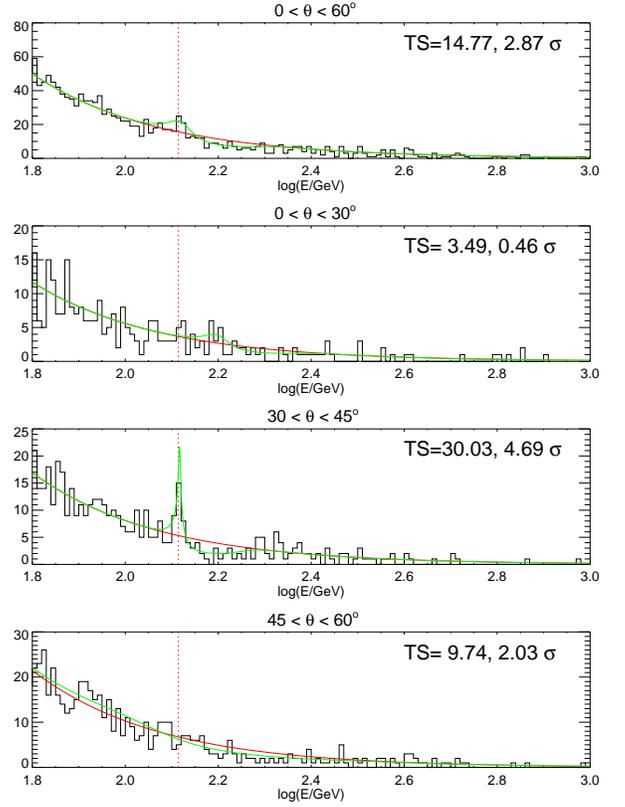}
  \caption{Fits of the energy mapping model to limb data for
  various ranges of inclination angle $\theta$.  The
  vertical red dotted line corresponds to 130 GeV.  The test
  statistic ($2\Delta\ln L$) for the best fit model (green
  line) relative to the null hypothesis (red line) is given,
  along with the significance, expressed in ``sigma''
  including a penalty for the 3 additional degrees of
  freedom.  The deviation from linearity is only significant
  in the $30\degree < \theta < 45\degree$ panel, but {\emph not} in events with other incidence angles.}
  \label{fig:limbfits}
\end{figure}

\begin{figure}
  \centering
  \includegraphics[width=0.45\textwidth]{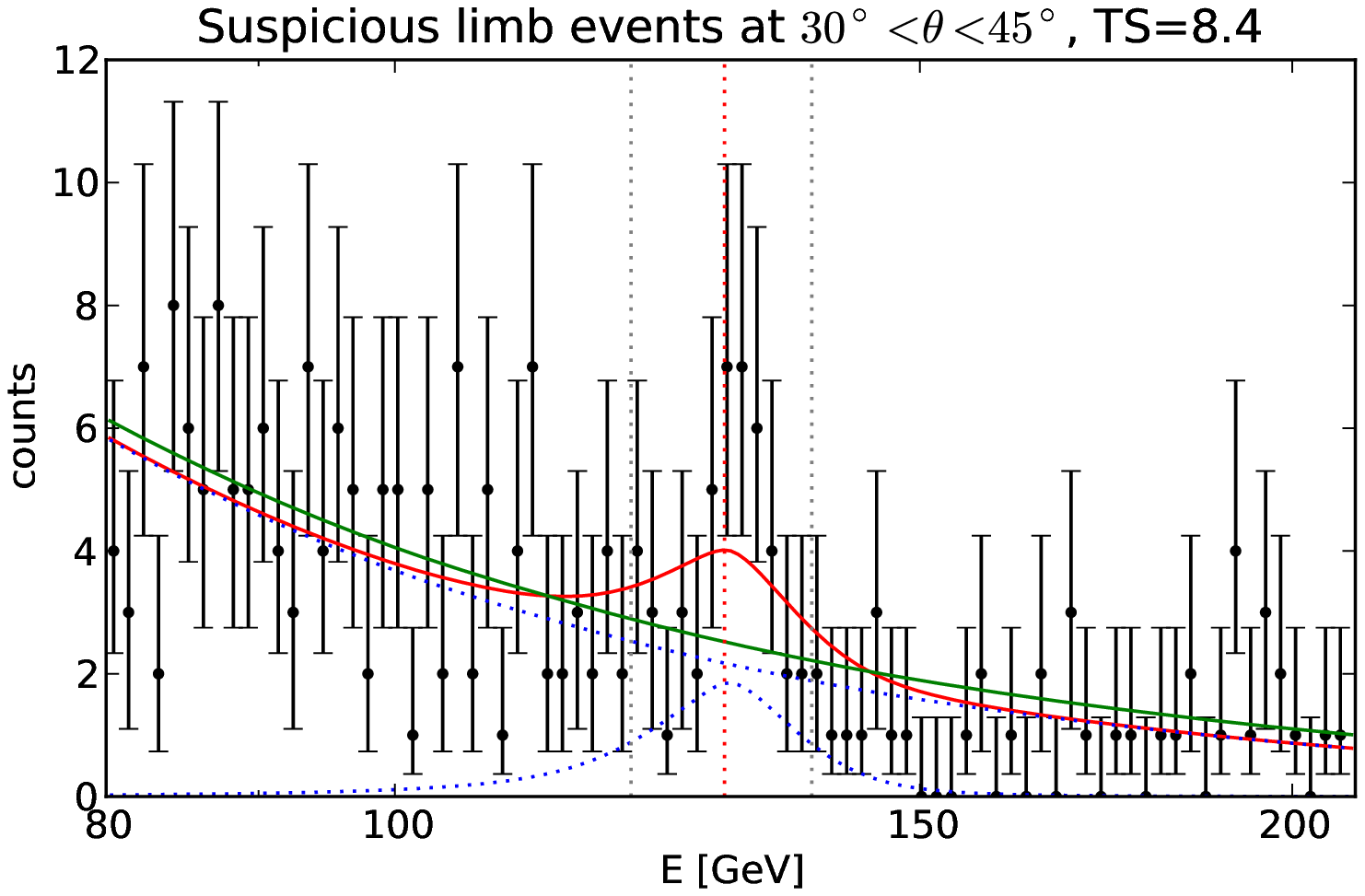}
  \includegraphics[width=0.45\textwidth]{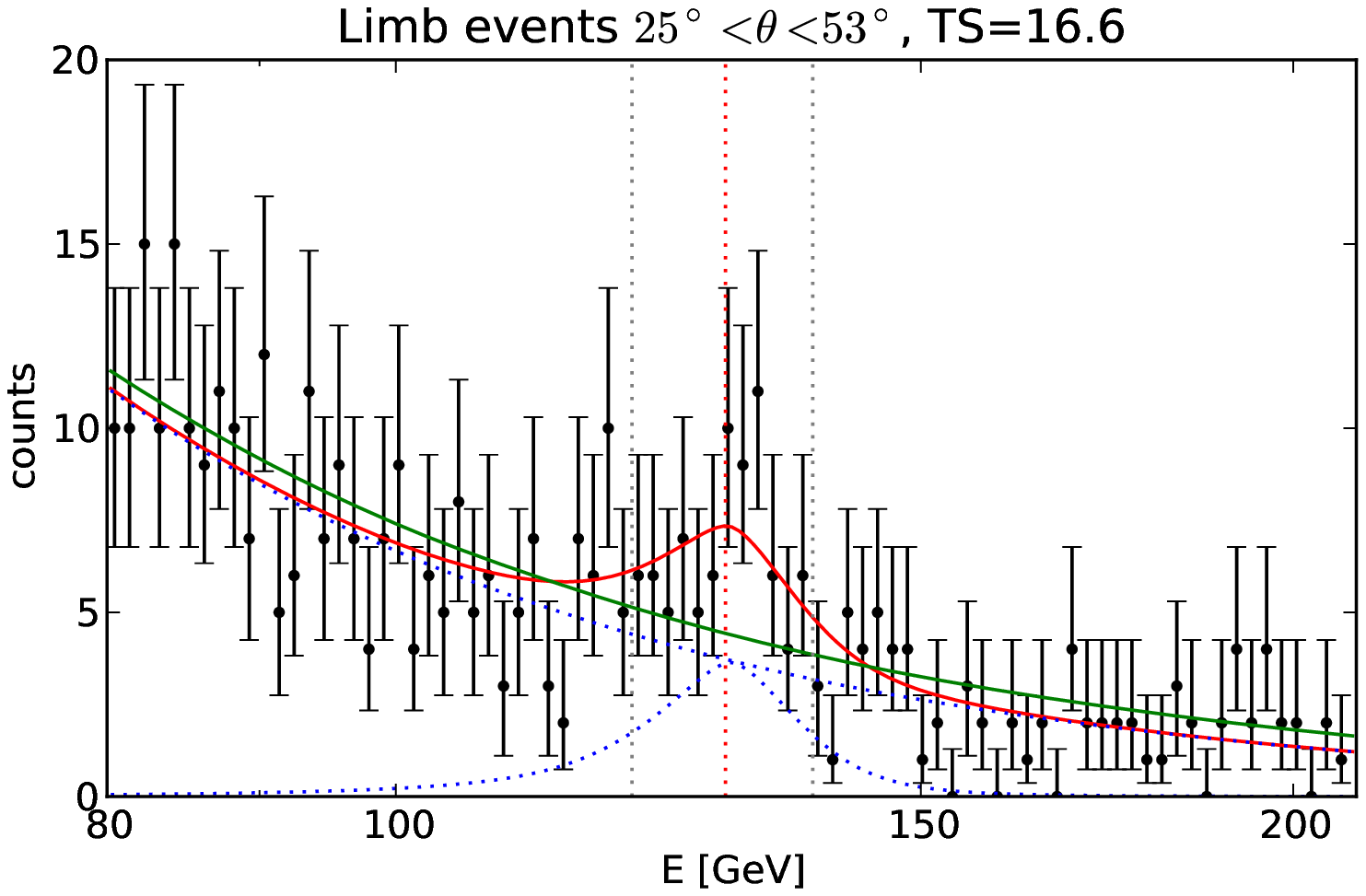}
  \includegraphics[width=0.45\textwidth]{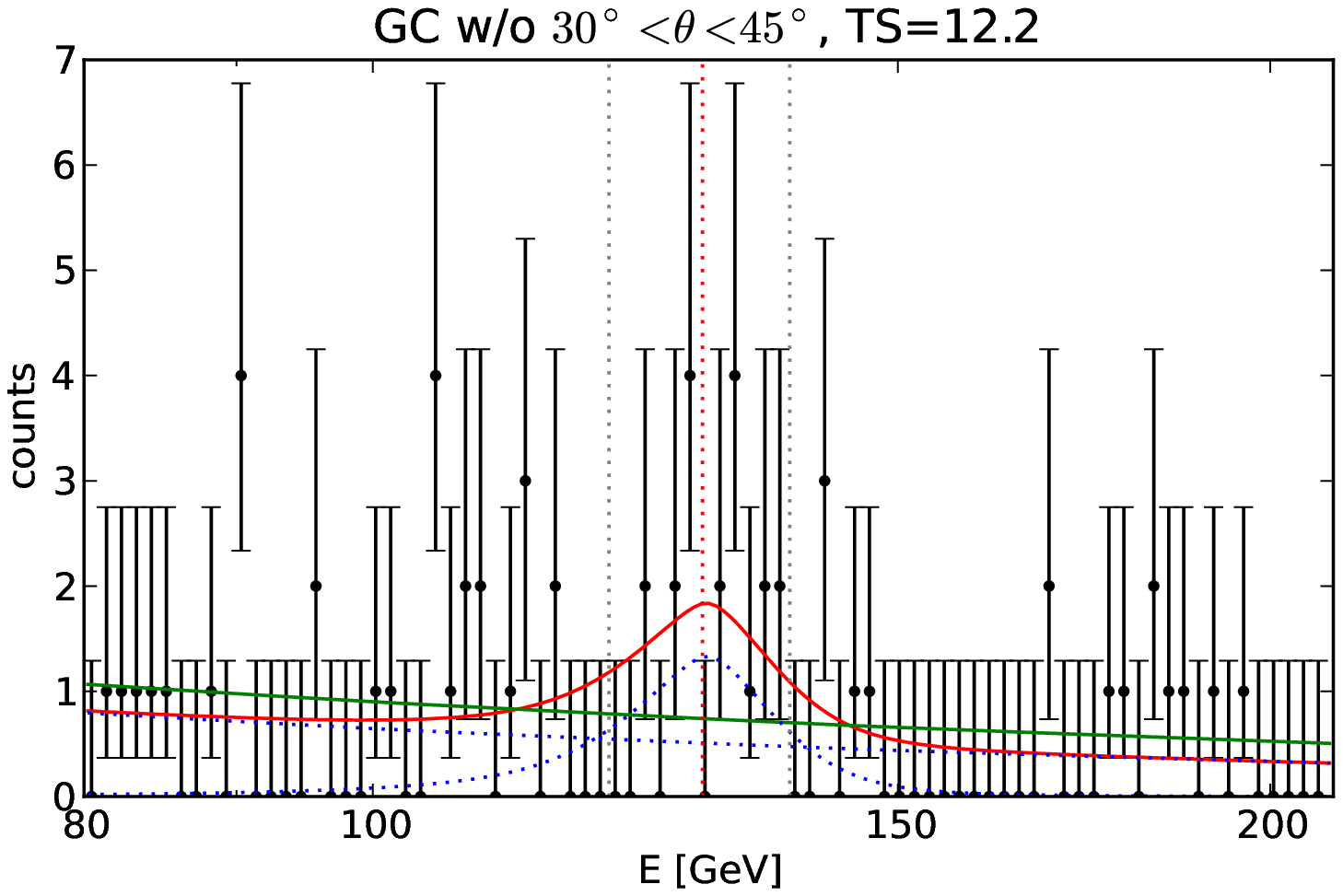}
  \caption{Same as Fig.~\ref{fig:spectra1}, but for
  suspicious limb data subset with
  $30^\circ<\theta<45^\circ$ (top panel),
  the $25^\circ<\theta<53^\circ$ subset tuned to give the largest $TS$ value (central panel),
  as well as for the
  GC events \emph{without} the problematic
  $30^\circ<\theta<45^\circ$ range for comparison (bottom panel).}
  \label{fig:spectra2}
\end{figure}

\begin{figure}
  \centering
  \includegraphics[width=1.0\linewidth]{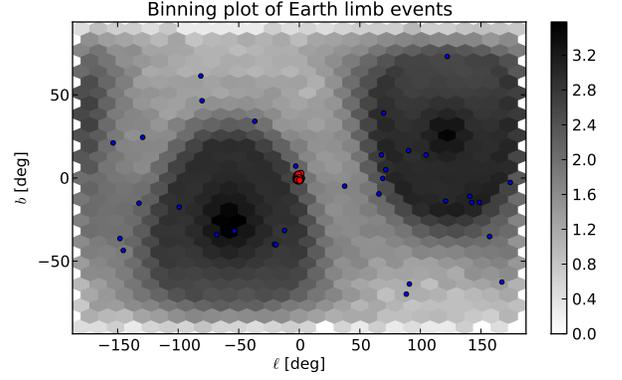}
  \caption{Earth limb events as function of Galactic coordinates $\ell$
  and $b$.  The majority of high-incidence limb events appear near the orbital
  pole, which precesses around the celestial pole. This pattern is expected
  from the observing strategy.  The GC line (\emph{red}) and Earth limb line
  (\emph{blue}) events are shown for comparison. The Earth limb line events do
  not originate in the Galactic center.}
  \label{fig:l-b}
\end{figure}

\begin{figure}
  \centering
  \includegraphics[width=0.45\textwidth]{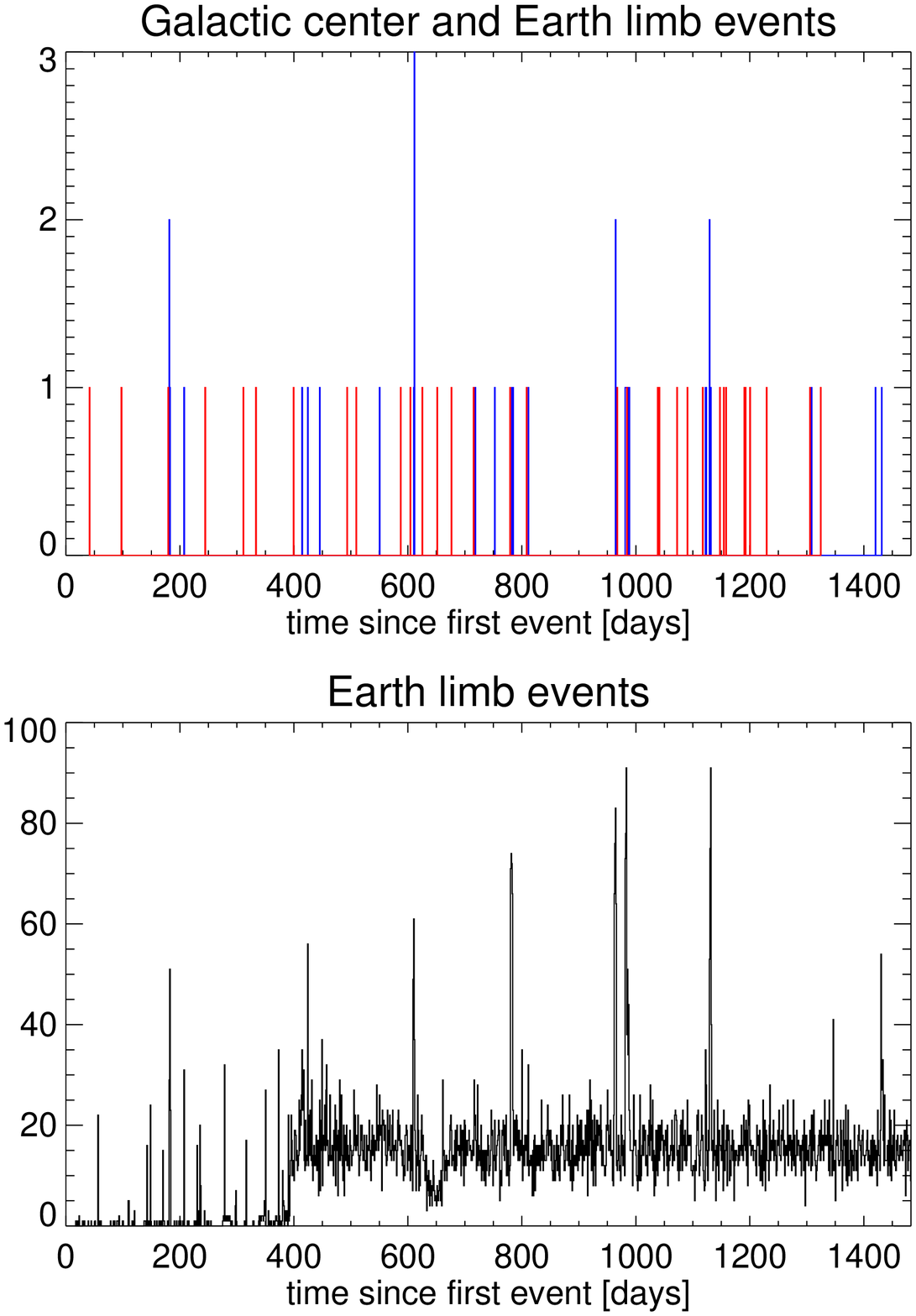}
  \caption{Time histogram (1-day bins) of GC line (\emph{red}) and Earth limb
    line (\emph{blue}) events (upper panel) and all limb events (lower panel).
    The Earth limb line events are only observed at high rocking angles that
    occur during occasional pointed observations. The survey mode rocking
    angle was changed from $35\degree$ to $50\degree$ at 400 days (lower
    panel).}
  \label{fig:timehist}
\end{figure}


In this section, we follow up the weak excess at $\sim$130 GeV 
in the limb photons
(Fig.~\ref{fig:polarPlotsAll})
and
characterize the nature of the apparent excess more precisely.  We show that
the critical incidence angles are not (solely) responsible for
the GC excess. 
We then
consider the possibility
that an energy mapping error could redistribute events in energy,
thus making a spectral feature, and
explore the various
parameters of the GC line events and the limb bump events.
\medskip

In Fig.~\ref{fig:theta-E-frontback}, we plot incidence angle $\theta$ versus
energy for limb events ($Z > 110\degree$).  For 125 GeV $< E <$ 135 GeV, we find
an excess of events with $30\degree < \theta < 45\degree$, as already indicated
in Fig.~\ref{fig:polarPlotsAll}.  This $\theta$ range contains 6\% of
the limb events for $E>100$ GeV. The 129 GeV bump is clearly visible in the
spectrum plotted in
the bottom panel of Fig.~\ref{fig:Ehist-all}, and decreases for
larger ranges of $\theta$ (other panels).  The bump appears equally in
\texttt{FRONT} and \texttt{BACK} converting events.

\subsection{Energy mapping error: a model for the limb bump}

 
As we discussed above in Sec.~II, it is very unlikely for the
GC line to be caused by extra events (from either photons with much higher
energies or CR background); the same is true for the line feature in the low
incidence Earth limb events.  Given these difficulties, we consider the
possibility that the Earth limb bump results from an energy mapping error, and
assume that for some unknown reason it only affects low incidence Earth limb
events (and potentially the GC).\footnote{Similar to anomalies in the
effective area, one would expect that such an error should affect all
regions of the sky, in contradiction to the observations.}

We propose a simple model, in
which the mapping from true energy to reported energy,
$E(E_t)$, is linear except for a bump near some reference
energy.  Smooth low-level perturbations over large energy
scales are not relevant here, and could be absorbed in the
effective area calibration.  In order to include a
small-scale bump in the response, we introduce a local
compact perturbation in the form of a Gaussian.  It is
convenient to work in logarithmic quantities, so we take
$x=\log E_t, y=\log E$, and \be
\label{eq:yofx}
y=x - A\sigma \exp\left(\frac{1}{2}-\frac{(x-x_0)^2}{2\sigma^2}\right),
\ee
where $A$ is a dimensionless amplitude of the bump ($-1<A<1$
is required for monotonicity of $y(x)$), $x_0$ is a
reference log energy, and $\sigma$ is the width of the bump (see
Fig.~\ref{fig:bumpmodel}).  The effect of the distortion is
to change the true spectrum $dN/dx = dN/dlog(E_t)$ into an
observed spectrum
\be
\label{eq:dndy}
\frac{dN}{dy} = \frac{dN}{dx} \left(\frac{dy}{dx}\right)^{-1} ,
\ee
with
\be
\label{eq:dydx}
\frac{dy}{dx} = 1 + A\sigma \exp\left(\frac{1}{2}-\frac{(x-x_0)^2}{2\sigma^2}\right)
\frac{x-x_0}{\sigma^2}.
\ee
Note that that the extreme values of $dy/dx = 1 \pm A$ occur
at $x-x_0 = \pm \sigma$ and at $y=x_0(\pm1-A)\sigma$.
Assuming the true limb spectrum is a power law, we may apply
this factor to obtain a model spectrum, and maximize the
Poisson likelihood of observing the data given the model.


In Fig.~\ref{fig:limbfits}, we fit the energy mapping model
to the Earth limb data for various ranges of inclination
angle. We find a 4.7$\sigma$ excess of $30\degree-45\degree$
limb photons at 129 GeV, with no significant excess at $0\degree - 30\degree$
or $45\degree - 60\degree$. 

\subsection{The Earth limb line and correlations with the GC signal}

As shown in the top panel of Fig.~\ref{fig:spectra2},
fitting the Earth limb events at
incidence angles $30^\circ<\theta<45^\circ$ with a
monochromatic line at 129 GeV instead of an energy remapping model yields a
local significance of only
$2.9\sigma$ (adopting an energy range from 80 to 210 GeV like above in
Fig.~\ref{fig:spectra1}).
However, a further tuning of the $\theta$-range yields significances up to $4.1\sigma$
(for $25^\circ<\theta<53^\circ$; central panel of Fig.~\ref{fig:spectra2}), 
but this comes with an additional number of trials.
In any case, the overall statistical
significance for a line in the $\theta<60^\circ$ Earth limb data is 
above $3\sigma$. 
For comparison, the bottom panel of Fig.~\ref{fig:spectra2} shows a fit to the Galactic center energy
spectrum \emph{without} incidence angles $30^\circ<\theta<45^\circ$. The GC
excess is not removed by this cut, which would have indicated a spurious
signal.  
Even when removing all events with
$25^\circ<\theta<53^\circ$ from the GC region (from region Reg4~\cite{Weniger:2012}),
we obtain $TS=5.1$ ($TS=10.1$) for the Galactic center signal, whereas the
Earth limb line completely disappears.

The Earth limb line events are distributed all over the sky, as expected
(Fig.~\ref{fig:l-b}).  The arrival time of these events is concentrated during
periods of high rocking angle, because it is geometrically impossible to see
limb
events at $\theta <45\degree$ in normal survey mode (Fig.~\ref{fig:timehist}).
As already mentioned in Section \ref{sec:130GeV}, the distribution of
$(\theta,\phi)$ vs.~each other and vs.~time and longitude are as expected
(Figs.~\ref{fig:phiThetaDist} and~\ref{fig:time_phi}).  In short, none of
these tests reveals suspicious trends or correlations, or indicates in any way that
a systematic error in detector coordinates could map events specifically onto
the Galactic center.


\section{Discussion and Conclusion}
\label{sec:Conclusion}

In this paper, we search the publicly available LAT data for any trends or
correlations that might indicate an instrumental origin for the spectral
feature at $\sim$130 GeV towards the GC \citep{Bringmann:2012, Weniger:2012, linepaper}. 
After addressing the following concerns, we find no evidence that the 130 GeV
feature is spurious. 
\medskip

\emph{The Galactic center is bright and has a hard spectrum.} On degree
scales, the GC surface brightness is less than a factor of 2 brighter than the
inner Galactic plane.  The inner plane provides an order of magnitude more
photons, and shows no sign of a 130 GeV bump.  Even larger samples (all limb
photons, all non-limb non-GC photons) also show no significant signal.  The GC has a hard
spectrum, and at TeV energies the
GC is brighter than the surrounding plane, but even if \emph{all} events above
300 GeV (assuming a hard spectrum $dN/dE\sim E^{-2}$) were incorrectly remapped
to 130 GeV, there would not be enough photons to explain the GC signal.

\emph{Observations of the GC have a restricted range of instrumental incidence
  angles.}  It is true that the survey strategy, orbital precession, and solar
panel alignment cause a non-trivial mapping of GC events onto $(\theta, \phi)$
as a function of time of year.  Specifically, they occur near the $x-z$ plane
($\phi=0$ or 180) when the Sun is near the GC or anti-center.  However, the GC
line events are drawn from the full range of $\theta$, $\phi$, and $t$.  The
limb line events are broadly distributed in $\theta$, $\phi$, and $\ell$, with
times corresponding to pointed observations at large zenith angle.  There is no
evidence that the distribution of line events deviates from expectations.

\emph{There are excess line events in the limb data for some incidence angles.}
For a small subset of the limb data with large rocking angle (when the limb may
be seen at small incidence angles) and a particular incidence angle range
around $30\degree$ to $45\degree$, we find a marginally significant 130 GeV feature
(above $3\sigma$).  
The majority of events with incidence angle $30\degree$ to
$45\degree$ are \emph{not} from the Earth limb, and we find no 130 GeV
feature in this much larger sample of events at these incidence angles.  If the
limb line events are an artifact, they must conspire to only appear when the
LAT is positioned at high rocking angle. 

\emph{The bump in the limb data might result from an energy mapping error.}
We propose a simple model for an error in the mapping from true photon energy
to reported energy.  This model reproduces the shape of the limb line feature
at 130 GeV and the dip at slightly higher energy, and has a local significance
of 4.7$\sigma$.  A modest amount of additional limb data would tell us if the
limb feature is a statistical fluke.  If the limb feature persists, it raises
serious concerns about the \texttt{Pass 7} processing of $E > 100$ GeV events.
\medskip

Additional limb data are available from the commissioning
period.  The Launch \& Early Operations (LEO) data were
taken during the first 60 days of the mission. Combined with
a dedicated Earth-limb observation in September 2008, this
provides $\sim$250 hours total livetime on the Earth
limb~\cite{FermiLimb}.

With \texttt{Pass 6} \texttt{diffuse} class events,
\cite{FermiLimb} has analyzed the spatial morphology and the
energy spectrum of the Earth limb sample,
which contains 218 photons above 100 GeV and 16 photons
above 500 GeV. The energy spectrum 
is a power-law with spectral index $2.79\pm
0.06$ for 3-500 GeV photons (Fig.~2 of \cite{FermiLimb}),
which is consistent with the primary cosmic-ray spectral
index $2.75\pm 0.03$.\footnote{The Earth limb photon above 10
  GeV from CR interactions in the upper atmospheric
  layers do not suffer large energy loses and the cosmic-ray
  primaries at this energy are unaffected by the Earth's
  magnetic fields. The Earth limb photons should have a
  spectral index close to that of the primary cosmic
  rays~\cite{FermiLimb}. } 

The spectrum of the Earth limb photons provided by
\citep{FermiLimb} does not show any significant feature at
130 GeV. If improved processing of the limb photons does not
replicate the line or the "energy mapping error" we found for a
subsample of the limb photons during the normal survey mode,
it can be dismissed as a statistical fluke.  If it
reappears, a deeper investigation into its cause will be necessary.

Even then, it is a challenge to understand how such an
instrumental feature could be mapped so precisely onto a
localized region within $5-10\degree$ of the GC.  The GC is
not near the path of the orbital pole, nor its axis of
precession.  The orbital phase, precession, Earth's orbit,
and time of year are all well mixed by the few $\times10^4$
orbits and 25 precession cycles over 1500 days.  We have
shown that the events in question are drawn from every part
of event and spacecraft parameter space available in the
public files. 

In summary, we find no significant instrumental systematics that could
plausibly explain the excess Galactic center emission observed at 130 GeV. 

\vskip 0.15in {\bf \noindent Note added:} During the final stages of this
work we became aware of another group discussing instrumental indications in
the Earth limb data~\cite{TempelSoon}.

\vskip 0.15in {\bf \noindent Acknowledgments:} We thank Neal
Weiner, Dan Hooper, and Jesse Thaler for helpful discussions. We acknowledge the use of
public data from the \Fermi\ data archive at
\texttt{http://fermi.gsfc.nasa.gov/ssc/}.  M.S. and
D.P.F. are partially supported by the NASA Fermi Guest
Investigator Program. Support for the work of M.S. was
provided by NASA through Einstein Postdoctoral Fellowship
grant number PF2-130102 awarded by the Chandra X-ray Center,
which is operated by the Smithsonian Astrophysical
Observatory for NASA under contract
NAS8-03060. C.W. acknowledges partial support from the
European 1231 Union FP7 ITN INVISIBLES (Marie Curie Actions,
PITN-GA-2011-289442).  This research made use of the NASA
Astrophysics Data System (ADS) and the IDL Astronomy User's
Library at Goddard (Available at
\texttt{http://idlastro.gsfc.nasa.gov}).

\section{Appendix}

\begin{figure}
  \centering
  \includegraphics[width=0.45\textwidth]{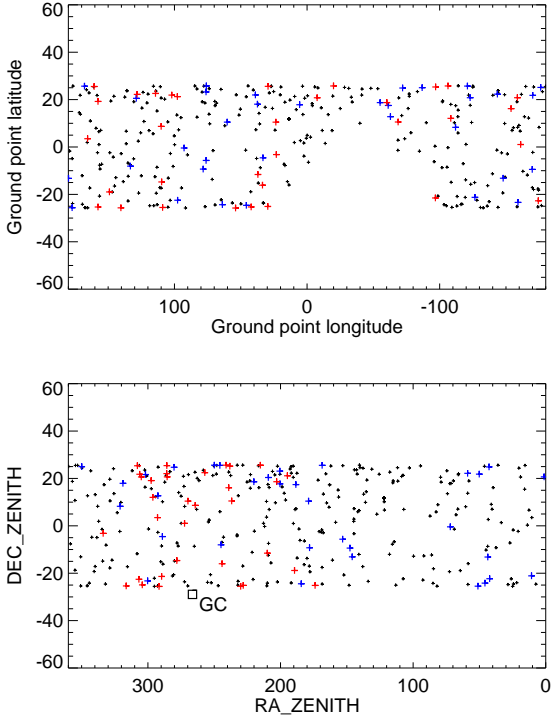}
  \caption{Distribution of events in Earth
  longitude and latitude (upper panel) and satellite zenith
  RA,dec (lower panel).  In this and following figures, Galactic center line photons (red), Earth limb line
  photons (blue) and Earth limb photons for all incidence angles (black) are
  shown.  The avoidance of the SAA leaves a
  hole in the upper panel.}
  \label{fig:geo-lonlat}
\end{figure}

\begin{figure}
  \centering
  \includegraphics[width=0.45\textwidth]{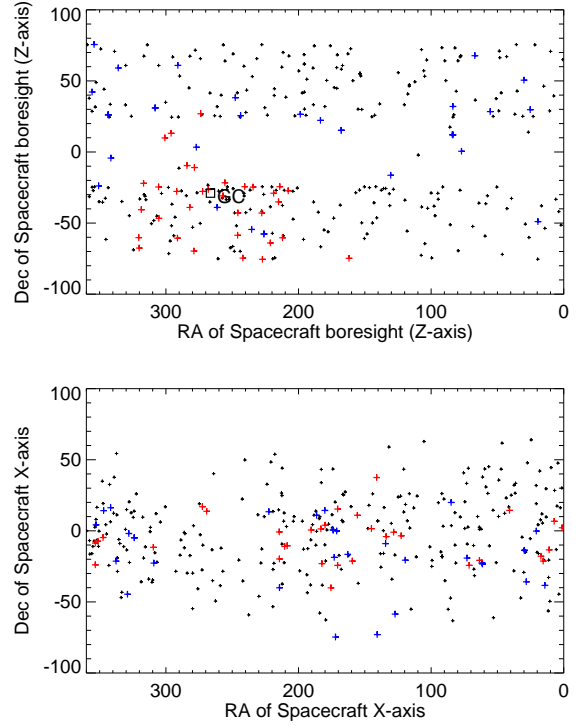}
  \caption{RA and Dec of spacecraft Z axis (boresight
  direction) and X axis (Solar panel direction).  }
  \label{fig:spacecraft-zx}
\end{figure}

\begin{figure}
  \centering
  \includegraphics[width=0.45\textwidth]{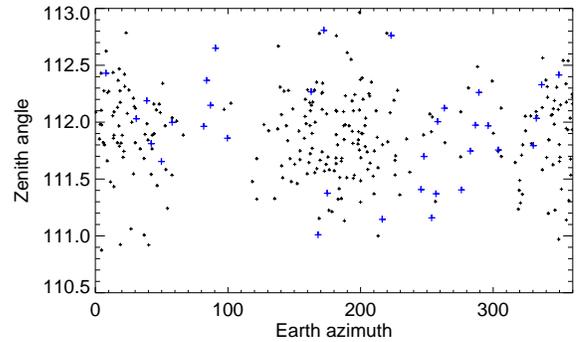}
  \caption{Zenith angle vs. Earth azimuth angle for limb
  photons.  As expected, the $\theta > 60\degree$ limb
  photons observed in survey mode are seen predominantly to
  the north and south azimuth directions, i.e approximately
  perpendicular to the orbit direction. The blue points are
  the Earth limb line events. }
  \label{fig:earth-az}
\end{figure}

\begin{figure}
  \centering
  \includegraphics[width=0.45\textwidth]{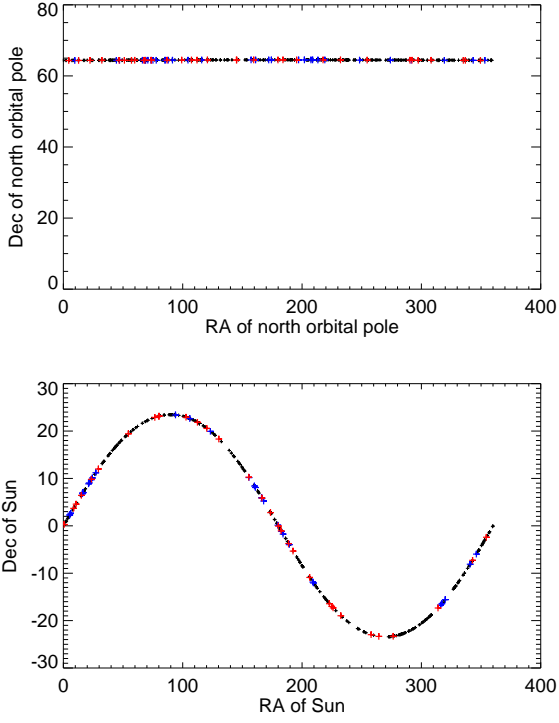}
  \caption{Events as a function of
  orbital precession phase (upper panel) and time of year
  (lower panel).  Note that there are only about 6 months of
  the year during which the 38 Earth limb line events (blue) are seen.}
  \label{fig:sun}
\end{figure}

\begin{figure}
  \centering
  \includegraphics[width=0.9\linewidth]{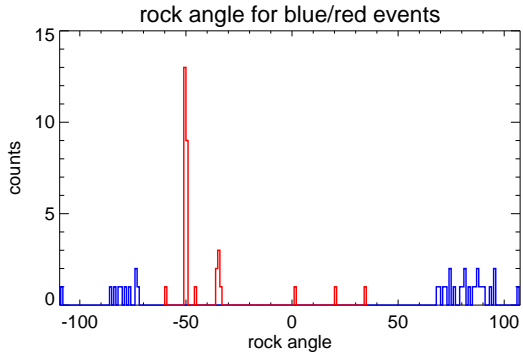}
  \caption{The rocking angle of the Galactic center line
  events (positive values indicate a rock angle toward the
  north from zenith.}
  \label{fig:rock}
\end{figure}

\begin{figure}
  \centering
  \includegraphics[width=0.9\linewidth]{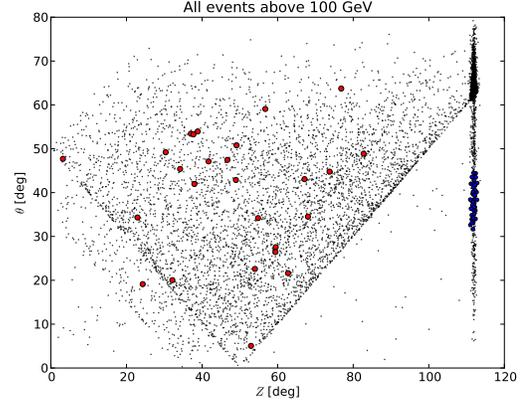}
  \caption{All events as function of zenith angle $Z$ and incidence angle
  $\theta$. The change of the rocking angle from $35^\circ$ to $50^\circ$ in
  2009 is clearly visible. None of the GC line events is observed near the
  horizon.}
  \label{fig:theta_z}
\end{figure}

In this Appendix, we compare the distribution of the
Galactic center line events with the Earth limb photons in
various projections of the event parameter space, and 
search for any unexpected
behavior.  As above, red points represent the Galactic
center line photons and the blue points represent
Earth limb line events. Earth limb
photons with 135 GeV $> E >$ 125 GeV at all incidence angles are shown in
black.  None of the Figures (Figs. \ref{fig:geo-lonlat}-\ref{fig:theta_z})
show any sign of unexpected behavior. 

\clearpage
\bibliography{systematics}

\end{document}